\documentclass[%
 aps,
 prl,
 superscriptaddress,
 amsmath,amssymb,
 reprint,%
]{revtex4-1}

\usepackage{ mathrsfs  }
\usepackage{graphicx}
\usepackage{dcolumn}
\usepackage{bm}
\usepackage[dvipsnames]{xcolor}

\usepackage{gensymb}

\newcommand{\gpc}{\dot\gamma_{c}}
\newcommand{\gpp}{\dot\gamma_{p}}
\newcommand{\gp}{\dot\gamma}

\newcommand{\Rcl}{R_\text{cl}}
\newcommand{\Rag}{R_\text{ag}}

\definecolor{color1}{rgb}{1.0000,1.0000,0.6250}
\definecolor{color2}{rgb}{1.0000,0.8958,0}
\definecolor{color3}{rgb}{1.0000,0.7396,0}
\definecolor{color5}{rgb}{1.0000,0.4792,0}
\definecolor{color7}{rgb}{1.0000,0.1146,0}
\definecolor{color10}{rgb}{0.75,0,0}
\definecolor{color15}{rgb}{0.3854,0,0}
\definecolor{color20}{rgb}{0,0.7294,0.6353}
\definecolor{color40}{rgb}{0,0.6902,0.6549}
\definecolor{color50}{rgb}{0,0.5333,0.7333}
\definecolor{color200}{rgb}{0,0.4157,0.7922}
\definecolor{color500}{rgb}{0,0,1}

\begin{document}

\title[Residual stresses and shear-induced overaging in boehmite gels]{Residual stresses and shear-induced overaging in boehmite gels
}

\author{Iana Sudreau\textsuperscript{*}}
\affiliation{IFP Energies nouvelles, Rond-point de l’échangeur de Solaize, BP 3, 69360 Solaize, France\looseness=-1}
\affiliation{ENSL, CNRS, Laboratoire de physique, F-69342 Lyon, France}
\author{Mathilde Auxois\textsuperscript{*}}
\affiliation{ENSL, CNRS, Laboratoire de physique, F-69342 Lyon, France}
\author{Marion Servel}%
\affiliation{IFP Energies nouvelles, Rond-point de l’échangeur de Solaize, BP 3, 69360 Solaize, France\looseness=-1}
\author{{\'Eric L\'ecolier}}%
\affiliation{{IFP Energies nouvelles, 1 \& 4, Avenue de Bois-Préau, 92852 Rueil-Malmaison, France\looseness=-1}}
\author{Sébastien Manneville}%
\affiliation{ENSL, CNRS, Laboratoire de physique, F-69342 Lyon, France}
\author{Thibaut Divoux}%
\affiliation{ENSL, CNRS, Laboratoire de physique, F-69342 Lyon, France}

\date{\today}

\begin{abstract}
Colloidal gels respond like soft solids at rest, whereas they flow like liquids under external shear. Starting from a fluidized state under an applied shear rate $\gpp$, abrupt flow cessation triggers a liquid-to-solid transition during which the stress relaxes towards a so-called \textit{residual stress} $\sigma_{\rm res}$ that tallies a macroscopic signature of previous shear history. Here, we report on the liquid-to-solid transition in gels of boehmite, an aluminum oxide, that shows a remarkable non-monotonic stress relaxation towards a residual stress $\sigma_{\rm res}(\gpp)$ characterized by a dual behavior relative to a critical value $\gpc$ of the shear rate $\gpp$. Following shear at $\gpp>\gpc$, the gel obtained upon flow cessation is insensitive to shear history, and the residual stress is negligible. However, for $\gpp<\gpc$, the gel encodes some memory of the shear history, and $\sigma_{\rm res}$ increases for decreasing shear rate, directly contributing to reinforcing the gel viscoelastic properties. Moreover, we show that both $\sigma_{\rm res}$ and the gel viscoelastic properties increase logarithmically with the strain accumulated during the shear period preceding flow cessation. Such a shear-induced ``overaging'' phenomenon bears great potential for tuning the rheological properties of colloidal gels.    
\end{abstract}

\maketitle

\textit{Introduction.-} Soft Glassy Materials (SGMs) are viscoelastic soft solids ubiquitous in science and engineering \cite{Balmforth:2014,Nelson:2019,Spicer:2020}. These materials display a shear-induced solid-to-liquid transition under large enough stresses, i.e., they flow like liquids above their yield stress $\sigma_y$, whereas SGMs recover their solidlike properties upon flow cessation \cite{Bonn:2017}. Such a liquid-to-amorphous-solid transition results in the partial relaxation of internal stresses, which usually follows a monotonic decay towards a so-called \textit{residual stress} $\sigma_{\rm res} < \sigma_y$ \cite{Keentok:1982,Nguyen:1992}. Residual stresses are critical to the mechanical properties of a broad range of materials across hard and soft condensed matter, for they control both the linear and non-linear material responses \cite{Withers:2001,Withers:2007}. 
In the case of SGMs, the residual stress encodes the sample memory of the plastic deformation accumulated prior to and during flow cessation \cite{Ballauff:2013,Vasisht:2021}. As such, $\sigma_{\rm res}$ depends on the shear rate $\gpp$ applied prior to flow cessation, and $\sigma_{\rm res}(\gpp)$ is usually reported to be a decreasing function, typically a weak power-law \cite{Osuji:2008,Negi:2009,Lidon:2017,Moghimi:2017b}, or a logarithm of $\gpp$ \cite{Mohan:2013} such that residual stresses become negligible following strong enough shear. 

From a microscopic point of view, the origin of the residual stress depends on the details of the SGMs microstructure. On the one hand, residual stresses in dense systems have been linked to the elastic contact forces in soft jammed particles \cite{Mohan:2013,Mohan:2013b,Mohan:2015}, and to the so-called ``supra-caging" effects in hard-sphere colloidal glasses \cite{Ballauff:2013}. On the other hand, residual stresses in dilute systems such as colloidal gels are associated with structural anisotropy frozen into the gel microstructure. Such anisotropy only emerges following low enough shear rates, i.e., low enough Peclet numbers, such that upon flow cessation, particles only move over short distances compared to the range of the inter-particle attraction, hence freezing their latest configuration \cite{Negi:2010,Moghimi:2017b}. Yet, the lack of information on residual stresses in colloidal gels compared to denser systems calls for more experiments. Moreover, while residual stresses and, more generally, memory effects in gels are often reported to be driven by the shear rate \cite{Koumakis:2015,Helal:2016,Jamali:2020}, little is known about the impact of the total strain accumulated during the shear period prior to flow cessation, and disentangling the contribution of the shear rate from that of the strain to residual stresses and to final viscoelastic properties remains challenging. 

In this Letter, we study the residual stress in acid-induced gels of boehmite, an aluminum oxide. We show that sudden flow cessation from a given shear rate $\gpp$  yields an anomalous non-monotonic stress relaxation towards a residual stress that strongly depends on the shear rate $\gpp$ applied before quenching the flow. We identify a critical shear rate $\gpc$, above which boehmite gels show negligible residual stresses, whereas, for $\gpp<\gpc$, they display significant residual stresses that increase when decreasing $\gpp$. Such a residual stress directly impacts the gel viscoelastic properties by reinforcing the elastic and viscous moduli, $G'$ and $G''$, respectively. In particular, for $\gpp<\gpc$, both $G'$ and $G''$ are directly proportional to the residual stress. Moreover, we show that aside from the shear rate $\gpp$, the total strain $\gamma_p$ accumulated before flow cessation plays a crucial role: the residual stress $\sigma_{\rm res}$ and the viscoelastic moduli grow logarithmically with $\gamma_p$, with a prefactor that depends on $\gpp$. Our results are prototypical of shear-induced ``overaging'', and show that boehmite gels can be tuned by both shear- and strain-controlled memory effects in the limit of low enough shear rates.     

\textit{Materials and methods.-} Steady shear and flow cessation experiments are performed on 4\%~vol.~boehmite gels prepared by dispersing a boehmite powder (Plural SB3, Sasol) in an aqueous solution of nitric acid at 14~g/L \cite{Ramsay.1978,Drouin.1988,Cristiani.2007}. After 35 min of mixing, the suspension contains unbreakable aggregates of diameter $2\Rag\simeq 100$~nm, which further assemble into large clusters of typical diameter $2 \Rcl\simeq 300$--1200~nm, leading to a sol-gel transition after a couple of hours \cite{Sudreau:2022}. The sample is left at rest for at least seven days prior to any rheological test, until the pH stabilizes to $\text{pH}=3.5$ \cite{Cristiani.2007,Fauchadour.2002}. About 10~mL of the gel sample is then transferred into a smooth cylindrical Couette cell of gap $1$~mm connected to a stress-controlled rheometer (AR-G2, TA Instruments).
\begin{figure}[t!]
    \centering
    \includegraphics[width=1\linewidth]{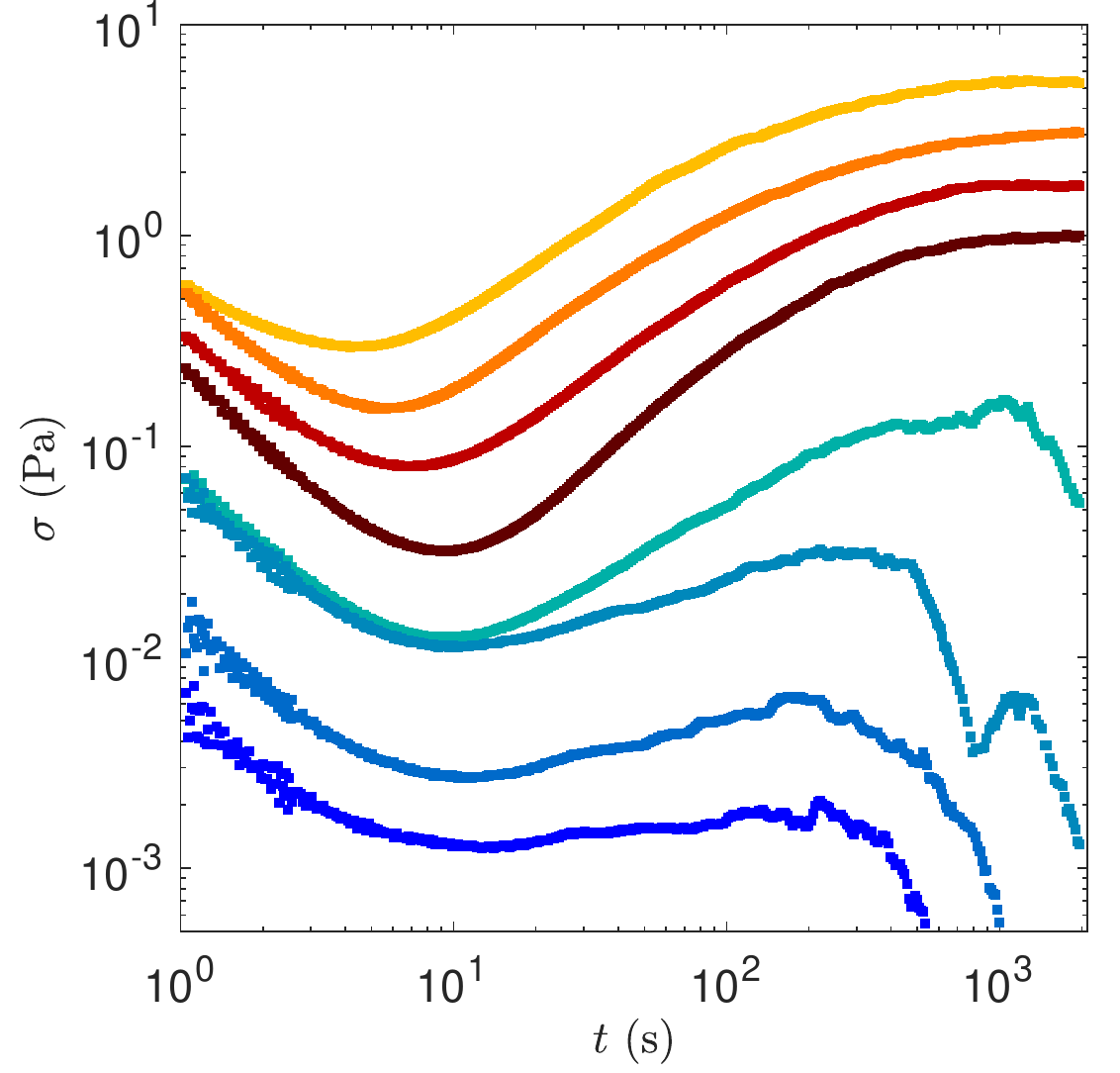}
    \caption{\label{Figure1} Stress response $\sigma(t)$ of a boehmite gel following flow cessation induced by imposing a decreasing step of shear rate from $\gp=\gpp$ to $\gp=0$ at $t=0$~s. {The target shear rate $\gp=0$ is reached in less than 1 s}. The colors code for $\gpp=3$~s$^{-1}$(\textcolor{color3}{$\blacksquare$}), 5~s$^{-1}$(\textcolor{color5}{$\blacksquare$}), 10~s$^{-1}$(\textcolor{color10}{$\blacksquare$}), 15~s$^{-1}$(\textcolor{color15}{$\blacksquare$}), 40~s$^{-1}$(\textcolor{color40}{$\blacksquare$}), 50~s$^{-1}$(\textcolor{color50}{$\blacksquare$}), 200~s$^{-1}$(\textcolor{color200}{$\blacksquare$}), and 500~s$^{-1}$(\textcolor{color500}{$\blacksquare$}). 
    }
\end{figure}

\textit{Results.-} The gel is first rejuvenated {by} a constant shear rate $\gp=1000$~s$^{-1}$ during 600~s {into a liquid suspension that displays no residual stress}. {Then} the flow is quenched to the shear rate of interest $\gpp$ for a duration $t_p=600$~s, { which is sufficient for the shear stress to reach a steady state whatever the value of $\gpp$ in the range of $1$~s$^{-1}$ to $1000$~s$^{-1}$}. Finally, the flow cessation, which sets the origin of time, is induced by imposing $\gp=0$~s$^{-1}$, while we monitor the stress relaxation $\sigma(t)$, as reported in Fig.~\ref{Figure1} for shear rates $\gpp$ ranging from $\gpp=3$~s$^{-1}$ to 500~s$^{-1}$. {For all $\gpp$ in the experimental range,} the stress displays a remarkable non-monotonic evolution, which becomes more pronounced for decreasing values of $\gpp$. More precisely, the stress response shows a dual behavior relative to a critical shear rate $\gpc\simeq 40$~s$^{-1}$. For $\gpp> \gpc$, the stress response first shows a linear decay until $t=t_{\rm min}\simeq 10$~s, followed by a stress overshoot, before decreasing again towards a negligible stress value, below the rheometer resolution. In stark contrast, upon flow cessation from a shear rate $\gpp<\gpc$, the stress response decreases down to a minimum value reached at time $t_{\rm min}\lesssim10$~s, and then increases towards a final and steady residual stress $\sigma_{\rm res}>0$ [see Fig.~S1 in Supplemental Material {for the temporal evolution of the shear rate $\gp$ effectively applied, and Fig.~S2} for longer stress relaxations and additional data recorded in a cone-and-plate geometry]. In that case, both  $t_{\rm min}$ and $\sigma_{\rm res}$ depend on the shear rate $\gpp$ applied before flow cessation.

\begin{figure}[t!]
    \centering
    \includegraphics[width=1\linewidth]{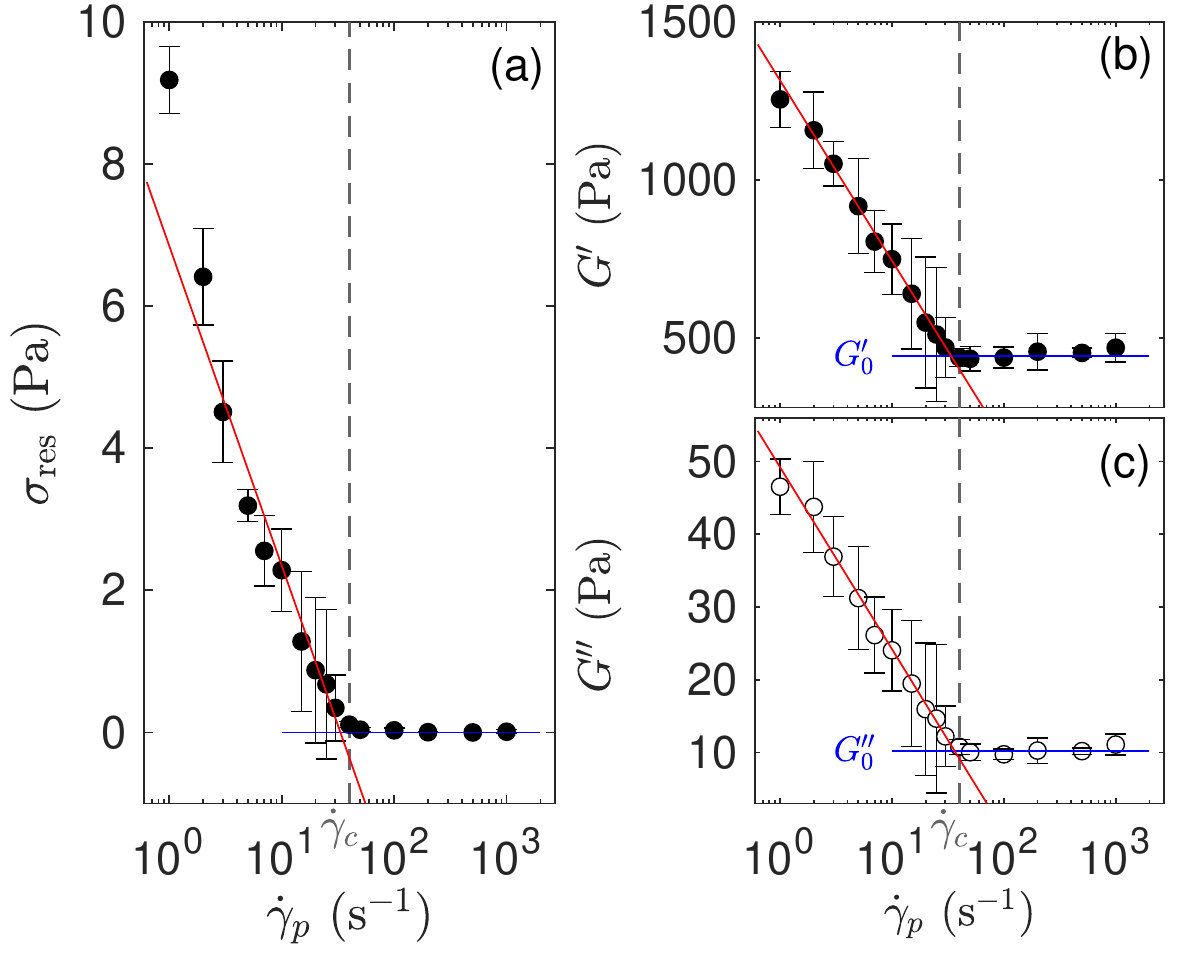}
    \caption{\label{Figure2} (a) Residual stress $\sigma_{\rm res}$, (b) elastic modulus $G'$, and (c) viscous modulus $G''$ vs. the shear rate $\dot \gamma_p$ applied prior to flow cessation. $\sigma_{\rm res}$ is measured at $t=2000$~s after flow cessation, while $G'$ and $G''$ are taken at $t=3000$~s. The vertical dashed line highlights the critical shear rate $\gpc= 40$~s$^{-1}$ below which the sample grows a residual stress, while $G'$ and $G''$ are reinforced. Blue lines highlight $\sigma_{\rm res}=0$~Pa, $G'_{0} = 444 \pm 35$~Pa, and $G''_{0} = 10 \pm 2$~Pa, and red lines are the best logarithmic fits of the data for $\gpp<\gpc$: $X=a_X\log\gpp+b_X$ with $a_\sigma=-4.5$ Pa, $a_{G'}=-570$ Pa, $a_{G''}=-25$~Pa, $b_\sigma=6.8$ Pa, $b_{G'}=1 315$ Pa, and $b_{G''}=49$ Pa. Error bars correspond to the standard deviations computed over 3 to 6 independent measurements.
    }
\end{figure}

\begin{figure*}[t!]
    \centering
    \includegraphics[width=1\linewidth]{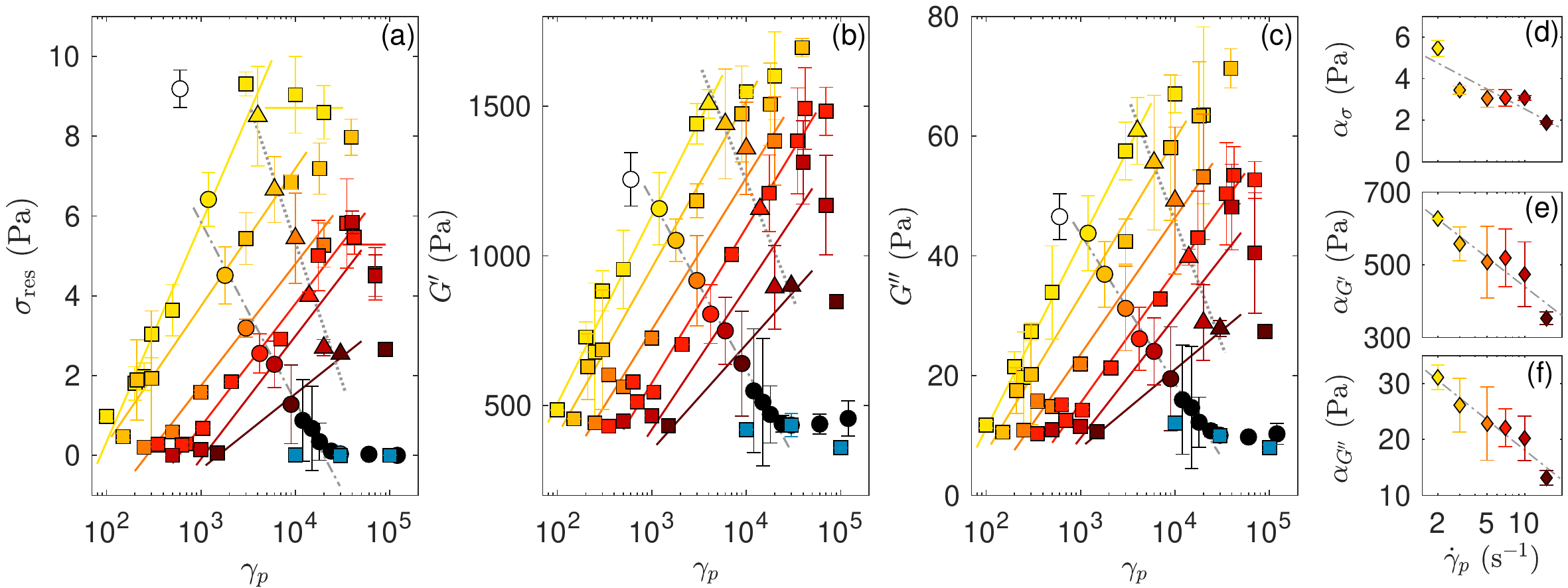}
    \caption{\label{Figure3}(a) Residual stress $\sigma_{\rm res}$, (b) elastic modulus $G'$, and (c) viscous modulus $G''$ vs.~the strain $\gamma_p$ accumulated prior to flow cessation. Colors code for $\gpp=1$~s$^{-1}$ ($\square$), 2~s$^{-1}$ (\textcolor{color2}{$\blacksquare$}), 3~s$^{-1}$(\textcolor{color3}{$\blacksquare$}), 5~s$^{-1}$(\textcolor{color5}{$\blacksquare$}), 7~s$^{-1}$(\textcolor{color7}{$\blacksquare$}), 10~s$^{-1}$(\textcolor{color10}{$\blacksquare$}), 15~s$^{-1}$(\textcolor{color15}{$\blacksquare$}), and 50~s$^{-1}$(\textcolor{color50}{$\blacksquare$}) imposed over various durations $t_p$ in independent experiments. Colored lines correspond to the best logarithmic fits [see Eq.~\eqref{eq1} in the main text]. Dashed and dotted gray lines highlight data measured for various shear rates $\gpp$ applied over a fixed duration $t_p=600$~s (circles, same data as in Fig.~\ref{Figure2}) and 2000~s (triangles), respectively. (d)--(f) slope $\alpha_X$ of the logarithmic increase with $\gamma_p$ of $X=\sigma_{\rm res}$, $G'$, and $G''$, respectively, vs.~$\gpp$. The gray dashed lines are guides to the eye.}
\end{figure*}

Figure~\ref{Figure2}(a) presents the residual stress $\sigma_{\rm res}$ measured 2000~s after flow cessation as a function of  $\gpp$ over three decades [see also Fig.~S3(a) in Supplemental Material for $t_{\rm min}(\gpp)$] and quantitatively confirms the dual behavior of boehmite gels: the residual stress is negligible for $\gpp>\gpc$, whereas it increases logarithmically with decreasing $\gpp$, up to $\sigma_{\rm res}\simeq 9$~Pa for $\gpp=1$~s$^{-1}$. These results show that boehmite gels display little if no memory of shear history for $\gpp>\gpc$, whereas they build up some frustration while quenching the gel from $\gpp<\gpc$, hence freezing residual stresses. {We further characterize the properties of these soft solids by measuring their steady-state linear viscoelastic properties at $t=3000$~s after flow cessation and at a single frequency of 1~Hz. Such a choice of frequency does not impact the generality of our results since the viscoelastic moduli of the present boehmite gels depend on the frequency as very weak power laws \cite{Sudreau:2022}}. Consistently with $\sigma_{\rm res}(\gpp)$, the elastic and viscous moduli, $G'$ and $G''$, both show a similar dual behavior [see also Fig.~S3(b) in Supplemental Material for $\tan\delta=G''/G'$]. Indeed, for $\gpp >\gpc$, the viscoelastic moduli are insensitive to shear history, with $G'_0=444\pm37$~Pa and $G''_0=10\pm2$~Pa. 
In contrast, for $\gpp<\gpc$, $G'$ and $G''$ increase logarithmically for decreasing $\gpp$,
which suggests that $\gpp$ affects both the skeletal backbone of the gel network, and its fractal microstructure, most likely through more anisotropy for smaller $\gpp$ \cite{Rooij:1994,Kao.2021}.
In summary, the shear rate $\gpp$ applied before flow cessation appears to control the terminal viscoelastic properties of boehmite gels. However, one should keep in mind that so far, the shearing duration was kept constant for all values of $\gpp$ so that the samples experienced different levels of strain prior to flow cessation. 

In order to disentangle the impact of the shear rate $\gpp$ from that of the accumulated strain $\gamma$ prior to flow cessation, we vary the shearing duration $t_p$ for given values of $\gpp$, so as to generate gels with different levels of accumulated strain $\gamma_p=\gpp t_p$ [see Fig.~S4 in Supplemental Material for corresponding stress relaxation data, {which all display a non-monotonic evolution}]. {Note that for $t_p<~$600 s, the stress does not reach a steady state for all levels of strain during the shearing step at $\gpp$.} Figure~\ref{Figure3} reveals the impact of both $\gamma_p$ and $\gpp$ on $\sigma_{\rm res}$, $G'$ and $G''$. For a fixed shear rate $\gpp<\gpc$, all three quantities increase logarithmically for increasing strain $\gamma_p$ [see colored lines in Fig.~\ref{Figure3}]. This result holds over three decades of strain for various shear rates. Moreover, the prefactors of this logarithmic dependence increase weakly for decreasing $\gpp$ [Fig.~\ref{Figure3}(d)--(f)], i.e., the lower the applied shear rate, the larger $\sigma_{\rm res}$, $G'$ and $G''$ for a fixed accumulated strain $\gamma_p$. In conclusion, the impact of both $\gpp$ and $\gamma_p$ on the various observables of interest, $X=\sigma_{\rm res}, G'$, or $G''$, is captured by: 
\begin{equation} \label{eq1}
\sigma_{\rm res},\, G',\, \mathrm{and}\, G'' = \alpha_X(\gpp)\log \gamma_p +\beta_X(\gpp),  
\end{equation}
where both $\alpha_X(\gpp)$ and $\beta_X(\gpp)$ are decreasing functions of $\gpp$. While the slope $\alpha_X(\gpp)$ depends only weakly on $\gpp$, the intercept $\beta_X(\gpp)$ accounts for the strong sensitivity on $\gpp$ reported in Fig.~\ref{Figure2}. Note that for $\gpp> \gpc$, neither the strain nor the shear rate have any impact on the residual stress or on the viscoelastic properties. 

\begin{figure}[t!]
    \centering
    \includegraphics[width=1\linewidth]{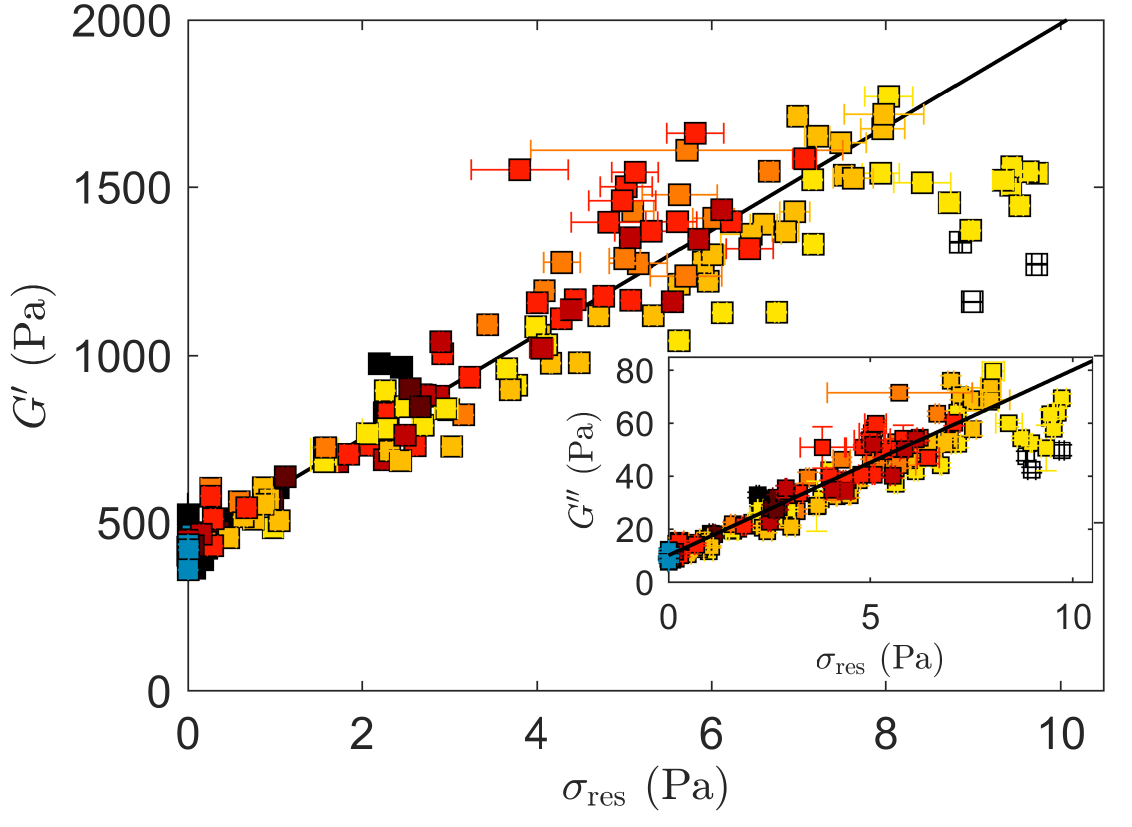}
    \caption{\label{Figure4} Elastic modulus $G'$ vs.~residual stress $\sigma_{\rm res}$ for a boehmite gel submitted to various shear histories, i.e., various shear rates $\gpp$ over different durations $t_p$. Data replotted from Fig.~\ref{Figure3}(a)--(c). Inset: Viscous modulus $G''$ vs.~$\sigma_{\rm res}$. Black lines are the best linear fits for $\sigma_{\rm res}<8.0$~Pa, respectively $G'=G'_0 (1+\lambda' \sigma_{\rm res})$ and $G''=G''_0 (1+\lambda'' \sigma_{\rm res})$, with $G'_0=444\pm37$~Pa, $\lambda'=0.35\pm0.04$, $G''_0=10\pm2$~Pa, and $\lambda''=0.69 \pm 0.06$.
    } 
\end{figure}

Finally, the robustness of Eq.~\eqref{eq1} prompts us to eliminate the strain and to directly link the gel viscoelastic properties to the residual stress. As pictured in Fig.~\ref{Figure4}, this allows us to collapse all data obtained for various shear histories, i.e., various shear rates $\gpp$ applied over different  durations $t_p$, hence yielding various strains $\gamma_p$, onto a master curve. Moreover, as expected from Eq.~\eqref{eq1}, the elastic and viscous moduli both increase proportionally to the residual stress, starting from the two reference values $G'_0$ and $G''_0$ measured for $\gpp>\gpc$ [see also Fig.~S5 in the Supplemental Material for the corresponding master curve for $\tan \delta$]. {Note that such a collapse does not depend on frequency (here 1~Hz), for viscoelastic properties of boehmite gels show a very weak power-law frequency dependence \cite{Sudreau:2022}.} Our results thus show that the viscoelastic properties of boehmite gels are reinforced by the residual stress, which grows due to an extended period of shear at low shear rates. 

\textit{Discussion and conclusion.-} Let us now summarize and discuss the most prominent results of this Letter. First, we have identified a critical shear rate $\gpc$ above which boehmite gels are insensitive to shear history. Such a critical shear rate is similar to the one reported in Ref.~\cite{Sudreau:2022}, although determined with a different rheological protocol. 
{In particular, as a key difference with Ref.~\cite{Sudreau:2022}, here, we fully rejuvenate the gel under $\gp=1000$~s$^{-1}$ before imposing the shear rate of interest $\dot \gamma_p$. Still, in both cases, a clear transition is observed in the evolution of the elastic modulus with $\gpp$ around the same values of $\gpc$ ($\gpc=30\pm 10$~s$^{-1}$ in Ref.~\cite{Sudreau:2022}).}
Our results thus confirm the robustness of $\gpc$, which was interpreted as a critical Mason number comparing the strength of shear to the bond force between two particles at contact \cite{Sudreau:2022,Markutsya:2014,Boromand:2017,Varga:2018,Varga:2019}. The hypothetical scenario proposed in Ref.~\cite{Sudreau:2022} to account for observations with $\gpp<\gpc$ is as follows: low enough shear rates do not fully rejuvenate the gel, which builds up some shear-induced structural anisotropy. Here, we show that for $\gpp<\gpc$, boehmite gels also develop internal stresses, which bear some memory of the flow history. 

Moreover, as reported in a broad variety of SGMs \cite{Derec:2003,Coussot:2006,Joshi.2018}, boehmite gels show spontaneous aging at rest characterized by a logarithmic time dependence of the viscoelastic moduli \cite{Sudreau:2022}. Such physical aging is a consequence of the system exploring its ``energy landscape'' by going from one metastable state to another under the effect of thermally activated dynamics \cite{Bouchaud:1992,Berthier:2003}. In this framework, moderate shear was shown to act as a delicate driving force that relocates the system into deeper energy minima, thus inducing ``overaging'' in SGMs \cite{Viasnoff:2002,Lacks:2004,Joshi:2014}. Here, the logarithmic increase of $G'$ and $G''$ with the strain accumulated under low enough shear rate is strikingly reminiscent of aging and constitutes a clear instance of ``overaging'' in a colloidal gel. Finally, we note that for vanishingly low shear rates, our results may be also linked to the so-called ``directed aging" recently achieved under a static load or strain in amorphous solids \cite{Pashine:2019,Hexner:2020}. 

Yet, the microscopic process underpinning shear-induced ``overaging'' remains to be identified in the present boehmite gels. It most likely originates either from local compaction, i.e., the slow increase in local volume fraction turning the local mechanical response from that of gel to that of a glass \cite{Johnson:2019,Keshavarz:2021}, or from the strain-induced alignment of the microstructure \cite{Mohraz:2005,Rajaram:2010,Santos:2013}. 
Interestingly such a strain-controlled reinforcement process seems to reach a limit for large enough strains. Indeed, when imposing  $\gpp=2$~s$^{-1}$ prior to flow cessation, the residual stress reaches a maximum at $\gamma_p^* \simeq 2.10^3$ before decreasing [see yellow symbols in Fig.~\ref{Figure3}(a)]. The same effect is visible beyond $\gamma_p^* \simeq 2.10^4$ for $\gpp=7$~s$^{-1}$ [see red symbols in Fig.~\ref{Figure3}(a)], suggesting that strain becomes detrimental beyond a certain point, breaking the strong microstructure that builds up at $\gamma_p<\gamma_p^*$. Such a scenario remains to be investigated with numerical tools, including molecular dynamics (MD) simulations, which would allow one to assess whether plasticity is responsible for the drop in residual stress beyond $\gamma_p^*$, and whether it results from rearrangements within gel strands \cite{Doorn:2018}, or occurs at larger length scales, between clusters \cite{Johnson:2018,Varga:2018}. 

Another signature of shear-induced ``overaging'' on the present boehmite gels for $\gpp<\gpc$ is their non-monotonic stress relaxation following flow cessation. Such an anomalous relaxation, common to disordered solids \cite{Murphy:2020,Mandal:2021}, has been recently reported in viscoelastic liquids and attributed to two key ingredients, namely the alignment of the microstructure under flow, and the formation of bonds upon flow cessation due to interparticle attractive forces that locally deform the bulk material, hence inducing internal stresses \cite{Hendricks:2019}. This picture based on shear-induced alignment is consistent with the emergence of some anisotropy in the microstructure of boehmite gels as hypothesized in Ref.~\cite{Sudreau:2022}, and with the growth of residual stresses only observed for $\gpp<\gpc$ in the present study. 
{In that framework, boehmite gels display a striking analogy with supramolecular systems, despite their terminal relaxation, for the non-monotonic response in these systems falls within the {physical network} part of their viscoelastic spectrum \cite{Hendricks:2019}.} Our results suggest that interparticle attractive forces are dominant for $\gpp<\gpc$, where residual stresses are observed. In contrast, hard-core repulsion takes over for $\gpp>\gpc$, accounting for the lack of residual stresses and the strong dampening of the non-monotonic stress relaxation. To our knowledge, non-monotonic relaxations in colloidal gels have not yet been captured by MD simulations \cite{Johnson:2019,Zia:2013}. Our data set thus constitutes a unique opportunity for future numerical work to identify the minimal ingredients necessary to reproduce such an anomalous relaxation. We anticipate that the anisotropy of boehmite particles and their noncentral interaction play a key role in the stress dynamics. 

{Finally, in cases where the stress relaxes to vanishingly small values, the late stages of the dynamics are consistently characterized by erratic fluctuations of $\sigma(t)$ [see Fig.~\ref{Figure1} and Fig.~S4 in the Supplemental Material]. Such fluctuations could be the signature of large-scale, heterogeneous dynamics occurring in the bulk or in the vicinity of the walls. Future experiments offering local insights into the sample microdynamics upon flow cessation should be undertaken to answer this open question.} 

{\textit{Conclusion.-}} Our study unravels a deep connection between residual stresses and viscoelastic properties in colloidal gels: the elastic and viscous modulus of boehmite gels are reinforced proportionally to the residual stress trapped upon flow cessation. Interestingly, such a linear relationship was previously reported in gels of fractal-like particles experiencing van der Waals attractive interactions \cite{Osuji:2008}. However, in contrast with the low shear rates explored here, the reinforcement in these fractal gels was observed in a shear-thickened state obtained only after a period of intense shearing, and following a monotonic stress relaxation. Such a discrepancy with our observations suggests that the linear dependence of the viscoelastic moduli with $\sigma_{\rm res}$ is somewhat universal in colloidal gels, regardless of microscopic details. As such, our experimental results constitute a benchmark for future modeling of residual stresses at the particle scale \cite{Bouzid:2020}, {not only to rationalize their relation to elasticity and reinforcement, but also to understand their dependence with the accumulated strain, here reported to be logarithmic, as well as their temporal evolution after flow cessation}.

\emph{Acknowledgments.-} We thank E.~Del Gado, T.~Gibaud, G.~Petekidis, and V.~Vasisht for fruitful discussions. 

\section*{References}

\begin{thebibliography}{54}%
\makeatletter
\providecommand \@ifxundefined [1]{%
 \@ifx{#1\undefined}
}%
\providecommand \@ifnum [1]{%
 \ifnum #1\expandafter \@firstoftwo
 \else \expandafter \@secondoftwo
 \fi
}%
\providecommand \@ifx [1]{%
 \ifx #1\expandafter \@firstoftwo
 \else \expandafter \@secondoftwo
 \fi
}%
\providecommand \natexlab [1]{#1}%
\providecommand \enquote  [1]{``#1''}%
\providecommand \bibnamefont  [1]{#1}%
\providecommand \bibfnamefont [1]{#1}%
\providecommand \citenamefont [1]{#1}%
\providecommand \href@noop [0]{\@secondoftwo}%
\providecommand \href [0]{\begingroup \@sanitize@url \@href}%
\providecommand \@href[1]{\@@startlink{#1}\@@href}%
\providecommand \@@href[1]{\endgroup#1\@@endlink}%
\providecommand \@sanitize@url [0]{\catcode `\\12\catcode `\$12\catcode
  `\&12\catcode `\#12\catcode `\^12\catcode `\_12\catcode `\%12\relax}%
\providecommand \@@startlink[1]{}%
\providecommand \@@endlink[0]{}%
\providecommand \url  [0]{\begingroup\@sanitize@url \@url }%
\providecommand \@url [1]{\endgroup\@href {#1}{\urlprefix }}%
\providecommand \urlprefix  [0]{URL }%
\providecommand \Eprint [0]{\href }%
\providecommand \doibase [0]{http://dx.doi.org/}%
\providecommand \selectlanguage [0]{\@gobble}%
\providecommand \bibinfo  [0]{\@secondoftwo}%
\providecommand \bibfield  [0]{\@secondoftwo}%
\providecommand \translation [1]{[#1]}%
\providecommand \BibitemOpen [0]{}%
\providecommand \bibitemStop [0]{}%
\providecommand \bibitemNoStop [0]{.\EOS\space}%
\providecommand \EOS [0]{\spacefactor3000\relax}%
\providecommand \BibitemShut  [1]{\csname bibitem#1\endcsname}%
\let\auto@bib@innerbib\@empty
\bibitem [{\citenamefont {Balmforth}\ \emph {et~al.}(2014)\citenamefont
  {Balmforth}, \citenamefont {Frigaard},\ and\ \citenamefont
  {Ovarlez}}]{Balmforth:2014}%
  \BibitemOpen
  \bibfield  {author} {\bibinfo {author} {\bibfnamefont {N.}~\bibnamefont
  {Balmforth}}, \bibinfo {author} {\bibfnamefont {I.}~\bibnamefont {Frigaard}},
  \ and\ \bibinfo {author} {\bibfnamefont {G.}~\bibnamefont {Ovarlez}},\
  }\href@noop {} {\bibfield  {journal} {\bibinfo  {journal} {Annu. Rev. Fluid
  Mech.}\ }\textbf {\bibinfo {volume} {46}},\ \bibinfo {pages} {121} (\bibinfo
  {year} {2014})}\BibitemShut {NoStop}%
\bibitem [{\citenamefont {Nelson}\ \emph {et~al.}(2019)\citenamefont {Nelson},
  \citenamefont {Schweizer}, \citenamefont {Rauzan}, \citenamefont {Nuzzo},
  \citenamefont {Vermant},\ and\ \citenamefont {Ewoldt}}]{Nelson:2019}%
  \BibitemOpen
  \bibfield  {author} {\bibinfo {author} {\bibfnamefont {A.~Z.}\ \bibnamefont
  {Nelson}}, \bibinfo {author} {\bibfnamefont {K.~S.}\ \bibnamefont
  {Schweizer}}, \bibinfo {author} {\bibfnamefont {B.~M.}\ \bibnamefont
  {Rauzan}}, \bibinfo {author} {\bibfnamefont {R.~G.}\ \bibnamefont {Nuzzo}},
  \bibinfo {author} {\bibfnamefont {J.}~\bibnamefont {Vermant}}, \ and\
  \bibinfo {author} {\bibfnamefont {R.~H.}\ \bibnamefont {Ewoldt}},\
  }\href@noop {} {\bibfield  {journal} {\bibinfo  {journal} {Curr. Opin. Solid
  State Mater. Sci.}\ }\textbf {\bibinfo {volume} {23}},\ \bibinfo {pages}
  {100758} (\bibinfo {year} {2019})}\BibitemShut {NoStop}%
\bibitem [{\citenamefont {Spicer}\ \emph {et~al.}(2020)\citenamefont {Spicer},
  \citenamefont {Caggioni},\ and\ \citenamefont {Squires}}]{Spicer:2020}%
  \BibitemOpen
  \bibfield  {author} {\bibinfo {author} {\bibfnamefont {P.~T.}\ \bibnamefont
  {Spicer}}, \bibinfo {author} {\bibfnamefont {M.}~\bibnamefont {Caggioni}}, \
  and\ \bibinfo {author} {\bibfnamefont {T.~M.}\ \bibnamefont {Squires}},\
  }\href@noop {} {\bibfield  {journal} {\bibinfo  {journal} {Chem. Eng. Prog.}\
  }\textbf {\bibinfo {volume} {116}},\ \bibinfo {pages} {32} (\bibinfo {year}
  {2020})}\BibitemShut {NoStop}%
\bibitem [{\citenamefont {Bonn}\ \emph {et~al.}(2017)\citenamefont {Bonn},
  \citenamefont {Denn}, \citenamefont {Berthier}, \citenamefont {Divoux},\ and\
  \citenamefont {Manneville}}]{Bonn:2017}%
  \BibitemOpen
  \bibfield  {author} {\bibinfo {author} {\bibfnamefont {D.}~\bibnamefont
  {Bonn}}, \bibinfo {author} {\bibfnamefont {M.~M.}\ \bibnamefont {Denn}},
  \bibinfo {author} {\bibfnamefont {L.}~\bibnamefont {Berthier}}, \bibinfo
  {author} {\bibfnamefont {T.}~\bibnamefont {Divoux}}, \ and\ \bibinfo {author}
  {\bibfnamefont {S.}~\bibnamefont {Manneville}},\ }\href@noop {} {\bibfield
  {journal} {\bibinfo  {journal} {Rev. Mod. Phys.}\ }\textbf {\bibinfo {volume}
  {89}} (\bibinfo {year} {2017})}\BibitemShut {NoStop}%
\bibitem [{\citenamefont {Keentok}(1982)}]{Keentok:1982}%
  \BibitemOpen
  \bibfield  {author} {\bibinfo {author} {\bibfnamefont {M.}~\bibnamefont
  {Keentok}},\ }\href@noop {} {\bibfield  {journal} {\bibinfo  {journal}
  {Rheol. Acta}\ }\textbf {\bibinfo {volume} {21}},\ \bibinfo {pages} {325}
  (\bibinfo {year} {1982})}\BibitemShut {NoStop}%
\bibitem [{\citenamefont {Nguyen}\ and\ \citenamefont
  {Boger}(1992)}]{Nguyen:1992}%
  \BibitemOpen
  \bibfield  {author} {\bibinfo {author} {\bibfnamefont {Q.}~\bibnamefont
  {Nguyen}}\ and\ \bibinfo {author} {\bibfnamefont {D.}~\bibnamefont {Boger}},\
  }\href@noop {} {\bibfield  {journal} {\bibinfo  {journal} {Annu. Rev. Fluid
  Mech.}\ }\textbf {\bibinfo {volume} {24}},\ \bibinfo {pages} {47} (\bibinfo
  {year} {1992})}\BibitemShut {NoStop}%
\bibitem [{\citenamefont {Withers}\ and\ \citenamefont
  {Bhadeshia}(2001)}]{Withers:2001}%
  \BibitemOpen
  \bibfield  {author} {\bibinfo {author} {\bibfnamefont {P.}~\bibnamefont
  {Withers}}\ and\ \bibinfo {author} {\bibfnamefont {H.}~\bibnamefont
  {Bhadeshia}},\ }\href {\doibase 10.1179/026708301101510087} {\bibfield
  {journal} {\bibinfo  {journal} {Mater. Sci. Tech.}\ }\textbf {\bibinfo
  {volume} {17}},\ \bibinfo {pages} {366} (\bibinfo {year} {2001})}\BibitemShut
  {NoStop}%
\bibitem [{\citenamefont {Withers}(2007)}]{Withers:2007}%
  \BibitemOpen
  \bibfield  {author} {\bibinfo {author} {\bibfnamefont {P.~J.}\ \bibnamefont
  {Withers}},\ }\href {https://doi.org/10.1088/0034-4885/70/12/r04} {\bibfield
  {journal} {\bibinfo  {journal} {Rep. Prog. Phys.}\ }\textbf {\bibinfo
  {volume} {70}},\ \bibinfo {pages} {2211} (\bibinfo {year}
  {2007})}\BibitemShut {NoStop}%
\bibitem [{\citenamefont {Ballauff}\ \emph {et~al.}(2013)\citenamefont
  {Ballauff}, \citenamefont {Brader}, \citenamefont {Egelhaaf}, \citenamefont
  {Fuchs}, \citenamefont {Horbach}, \citenamefont {Koumakis}, \citenamefont
  {Kr{\"u}ger}, \citenamefont {Laurati}, \citenamefont {Mutch}, \citenamefont
  {Petekidis}, \citenamefont {Siebenburger}, \citenamefont {Voigtmann},\ and\
  \citenamefont {Zausch}}]{Ballauff:2013}%
  \BibitemOpen
  \bibfield  {author} {\bibinfo {author} {\bibfnamefont {M.}~\bibnamefont
  {Ballauff}}, \bibinfo {author} {\bibfnamefont {J.~M.}\ \bibnamefont
  {Brader}}, \bibinfo {author} {\bibfnamefont {S.~U.}\ \bibnamefont
  {Egelhaaf}}, \bibinfo {author} {\bibfnamefont {M.}~\bibnamefont {Fuchs}},
  \bibinfo {author} {\bibfnamefont {J.}~\bibnamefont {Horbach}}, \bibinfo
  {author} {\bibfnamefont {N.}~\bibnamefont {Koumakis}}, \bibinfo {author}
  {\bibfnamefont {M.}~\bibnamefont {Kr{\"u}ger}}, \bibinfo {author}
  {\bibfnamefont {M.}~\bibnamefont {Laurati}}, \bibinfo {author} {\bibfnamefont
  {K.~J.}\ \bibnamefont {Mutch}}, \bibinfo {author} {\bibfnamefont
  {G.}~\bibnamefont {Petekidis}}, \bibinfo {author} {\bibfnamefont
  {M.}~\bibnamefont {Siebenburger}}, \bibinfo {author} {\bibfnamefont
  {T.}~\bibnamefont {Voigtmann}}, \ and\ \bibinfo {author} {\bibfnamefont
  {J.}~\bibnamefont {Zausch}},\ }\href@noop {} {\bibfield  {journal} {\bibinfo
  {journal} {Phys. Rev. Lett.}\ }\textbf {\bibinfo {volume} {110}},\ \bibinfo
  {pages} {215701} (\bibinfo {year} {2013})}\BibitemShut {NoStop}%
\bibitem [{\citenamefont {Vasisht}\ \emph {et~al.}(2021)\citenamefont
  {Vasisht}, \citenamefont {Chaudhuri},\ and\ \citenamefont
  {Martens}}]{Vasisht:2021}%
  \BibitemOpen
  \bibfield  {author} {\bibinfo {author} {\bibfnamefont {V.~V.}\ \bibnamefont
  {Vasisht}}, \bibinfo {author} {\bibfnamefont {P.}~\bibnamefont {Chaudhuri}},
  \ and\ \bibinfo {author} {\bibfnamefont {K.}~\bibnamefont {Martens}},\
  }\href@noop {} {\bibfield  {journal} {\bibinfo  {journal} {arXiv:2108.12782}\
  } (\bibinfo {year} {2021})}\BibitemShut {NoStop}%
\bibitem [{\citenamefont {Osuji}\ \emph {et~al.}(2008)\citenamefont {Osuji},
  \citenamefont {Kim},\ and\ \citenamefont {Weitz}}]{Osuji:2008}%
  \BibitemOpen
  \bibfield  {author} {\bibinfo {author} {\bibfnamefont {C.~O.}\ \bibnamefont
  {Osuji}}, \bibinfo {author} {\bibfnamefont {C.}~\bibnamefont {Kim}}, \ and\
  \bibinfo {author} {\bibfnamefont {D.~A.}\ \bibnamefont {Weitz}},\ }\href@noop
  {} {\bibfield  {journal} {\bibinfo  {journal} {Phys. Rev. E}\ }\textbf
  {\bibinfo {volume} {77}},\ \bibinfo {pages} {060402(R)} (\bibinfo {year}
  {2008})}\BibitemShut {NoStop}%
\bibitem [{\citenamefont {Negi}\ and\ \citenamefont {Osuji}(2009)}]{Negi:2009}%
  \BibitemOpen
  \bibfield  {author} {\bibinfo {author} {\bibfnamefont {A.~S.}\ \bibnamefont
  {Negi}}\ and\ \bibinfo {author} {\bibfnamefont {C.~O.}\ \bibnamefont
  {Osuji}},\ }\href@noop {} {\bibfield  {journal} {\bibinfo  {journal} {Phys.
  Rev. E}\ }\textbf {\bibinfo {volume} {80}},\ \bibinfo {pages} {010404}
  (\bibinfo {year} {2009})}\BibitemShut {NoStop}%
\bibitem [{\citenamefont {Lidon}\ \emph {et~al.}(2017)\citenamefont {Lidon},
  \citenamefont {Villa},\ and\ \citenamefont {Manneville}}]{Lidon:2017}%
  \BibitemOpen
  \bibfield  {author} {\bibinfo {author} {\bibfnamefont {P.}~\bibnamefont
  {Lidon}}, \bibinfo {author} {\bibfnamefont {L.}~\bibnamefont {Villa}}, \ and\
  \bibinfo {author} {\bibfnamefont {S.}~\bibnamefont {Manneville}},\ }\href
  {\doibase 10.1007/s00397-016-0961-4} {\bibfield  {journal} {\bibinfo
  {journal} {Rheol. Acta.}\ }\textbf {\bibinfo {volume} {56}},\ \bibinfo
  {pages} {307} (\bibinfo {year} {2017})}\BibitemShut {NoStop}%
\bibitem [{\citenamefont {Moghimi}\ \emph {et~al.}(2017)\citenamefont
  {Moghimi}, \citenamefont {Jacob},\ and\ \citenamefont
  {Petekidis}}]{Moghimi:2017b}%
  \BibitemOpen
  \bibfield  {author} {\bibinfo {author} {\bibfnamefont {E.}~\bibnamefont
  {Moghimi}}, \bibinfo {author} {\bibfnamefont {A.~R.}\ \bibnamefont {Jacob}},
  \ and\ \bibinfo {author} {\bibfnamefont {G.}~\bibnamefont {Petekidis}},\
  }\href {\doibase 10.1039/C7SM01655G} {\bibfield  {journal} {\bibinfo
  {journal} {Soft Matter}\ }\textbf {\bibinfo {volume} {13}},\ \bibinfo {pages}
  {7824} (\bibinfo {year} {2017})}\BibitemShut {NoStop}%
\bibitem [{\citenamefont {Mohan}\ \emph
  {et~al.}(2013{\natexlab{a}})\citenamefont {Mohan}, \citenamefont
  {Bonnecaze},\ and\ \citenamefont {Cloitre}}]{Mohan:2013}%
  \BibitemOpen
  \bibfield  {author} {\bibinfo {author} {\bibfnamefont {L.}~\bibnamefont
  {Mohan}}, \bibinfo {author} {\bibfnamefont {R.}~\bibnamefont {Bonnecaze}}, \
  and\ \bibinfo {author} {\bibfnamefont {M.}~\bibnamefont {Cloitre}},\
  }\href@noop {} {\bibfield  {journal} {\bibinfo  {journal} {Phys. Rev. Lett.}\
  }\textbf {\bibinfo {volume} {111}},\ \bibinfo {pages} {268301} (\bibinfo
  {year} {2013}{\natexlab{a}})}\BibitemShut {NoStop}%
\bibitem [{\citenamefont {Mohan}\ \emph
  {et~al.}(2013{\natexlab{b}})\citenamefont {Mohan}, \citenamefont {Pellet},
  \citenamefont {Cloitre},\ and\ \citenamefont {Bonnecaze}}]{Mohan:2013b}%
  \BibitemOpen
  \bibfield  {author} {\bibinfo {author} {\bibfnamefont {L.}~\bibnamefont
  {Mohan}}, \bibinfo {author} {\bibfnamefont {C.}~\bibnamefont {Pellet}},
  \bibinfo {author} {\bibfnamefont {M.}~\bibnamefont {Cloitre}}, \ and\
  \bibinfo {author} {\bibfnamefont {R.}~\bibnamefont {Bonnecaze}},\ }\href@noop
  {} {\bibfield  {journal} {\bibinfo  {journal} {J. Rheol.}\ }\textbf {\bibinfo
  {volume} {57}},\ \bibinfo {pages} {1023} (\bibinfo {year}
  {2013}{\natexlab{b}})}\BibitemShut {NoStop}%
\bibitem [{\citenamefont {Mohan}\ \emph {et~al.}(2015)\citenamefont {Mohan},
  \citenamefont {Cloitre},\ and\ \citenamefont {Bonnecaze}}]{Mohan:2015}%
  \BibitemOpen
  \bibfield  {author} {\bibinfo {author} {\bibfnamefont {L.}~\bibnamefont
  {Mohan}}, \bibinfo {author} {\bibfnamefont {M.}~\bibnamefont {Cloitre}}, \
  and\ \bibinfo {author} {\bibfnamefont {R.~T.}\ \bibnamefont {Bonnecaze}},\
  }\href {\doibase 10.1122/1.4901750} {\bibfield  {journal} {\bibinfo
  {journal} {J. Rheol.}\ }\textbf {\bibinfo {volume} {59}},\ \bibinfo {pages}
  {63} (\bibinfo {year} {2015})}\BibitemShut {NoStop}%
\bibitem [{\citenamefont {Negi}\ and\ \citenamefont {Osuji}(2010)}]{Negi:2010}%
  \BibitemOpen
  \bibfield  {author} {\bibinfo {author} {\bibfnamefont {A.~S.}\ \bibnamefont
  {Negi}}\ and\ \bibinfo {author} {\bibfnamefont {C.~O.}\ \bibnamefont
  {Osuji}},\ }\href@noop {} {\bibfield  {journal} {\bibinfo  {journal} {J.
  Rheol.}\ }\textbf {\bibinfo {volume} {54}},\ \bibinfo {pages} {943} (\bibinfo
  {year} {2010})}\BibitemShut {NoStop}%
\bibitem [{\citenamefont {Koumakis}\ \emph {et~al.}(2015)\citenamefont
  {Koumakis}, \citenamefont {Moghimi}, \citenamefont {Besseling}, \citenamefont
  {Poon}, \citenamefont {Brady},\ and\ \citenamefont
  {Petekidis}}]{Koumakis:2015}%
  \BibitemOpen
  \bibfield  {author} {\bibinfo {author} {\bibfnamefont {N.}~\bibnamefont
  {Koumakis}}, \bibinfo {author} {\bibfnamefont {E.}~\bibnamefont {Moghimi}},
  \bibinfo {author} {\bibfnamefont {R.}~\bibnamefont {Besseling}}, \bibinfo
  {author} {\bibfnamefont {W.~C.~K.}\ \bibnamefont {Poon}}, \bibinfo {author}
  {\bibfnamefont {J.~F.}\ \bibnamefont {Brady}}, \ and\ \bibinfo {author}
  {\bibfnamefont {G.}~\bibnamefont {Petekidis}},\ }\href@noop {} {\bibfield
  {journal} {\bibinfo  {journal} {Soft Matter}\ }\textbf {\bibinfo {volume}
  {11}},\ \bibinfo {pages} {4640} (\bibinfo {year} {2015})}\BibitemShut
  {NoStop}%
\bibitem [{\citenamefont {Helal}\ \emph {et~al.}(2016)\citenamefont {Helal},
  \citenamefont {Divoux},\ and\ \citenamefont {McKinley}}]{Helal:2016}%
  \BibitemOpen
  \bibfield  {author} {\bibinfo {author} {\bibfnamefont {A.}~\bibnamefont
  {Helal}}, \bibinfo {author} {\bibfnamefont {T.}~\bibnamefont {Divoux}}, \
  and\ \bibinfo {author} {\bibfnamefont {G.~H.}\ \bibnamefont {McKinley}},\
  }\href@noop {} {\bibfield  {journal} {\bibinfo  {journal} {Phys. Rev.
  Applied}\ }\textbf {\bibinfo {volume} {6}},\ \bibinfo {pages} {064004}
  (\bibinfo {year} {2016})}\BibitemShut {NoStop}%
\bibitem [{\citenamefont {Jamali}\ \emph {et~al.}(2020)\citenamefont {Jamali},
  \citenamefont {Armstrong},\ and\ \citenamefont {McKinley}}]{Jamali:2020}%
  \BibitemOpen
  \bibfield  {author} {\bibinfo {author} {\bibfnamefont {S.}~\bibnamefont
  {Jamali}}, \bibinfo {author} {\bibfnamefont {R.~C.}\ \bibnamefont
  {Armstrong}}, \ and\ \bibinfo {author} {\bibfnamefont {G.~H.}\ \bibnamefont
  {McKinley}},\ }\href@noop {} {\bibfield  {journal} {\bibinfo  {journal}
  {Mater. Today Adv.}\ }\textbf {\bibinfo {volume} {5}},\ \bibinfo {pages}
  {100026} (\bibinfo {year} {2020})}\BibitemShut {NoStop}%
\bibitem [{\citenamefont {Ramsay}\ \emph {et~al.}(1978)\citenamefont {Ramsay},
  \citenamefont {Daish},\ and\ \citenamefont {Wright}}]{Ramsay.1978}%
  \BibitemOpen
  \bibfield  {author} {\bibinfo {author} {\bibfnamefont {J.~D.~F.}\
  \bibnamefont {Ramsay}}, \bibinfo {author} {\bibfnamefont {S.~R.}\
  \bibnamefont {Daish}}, \ and\ \bibinfo {author} {\bibfnamefont {C.~J.}\
  \bibnamefont {Wright}},\ }\href@noop {} {\bibfield  {journal} {\bibinfo
  {journal} {Faraday Discuss.}\ }\textbf {\bibinfo {volume} {65}},\ \bibinfo
  {pages} {65} (\bibinfo {year} {1978})}\BibitemShut {NoStop}%
\bibitem [{\citenamefont {Drouin}\ \emph {et~al.}(1988)\citenamefont {Drouin},
  \citenamefont {Chopin}, \citenamefont {Nortier},\ and\ \citenamefont {{Van
  Damme}}}]{Drouin.1988}%
  \BibitemOpen
  \bibfield  {author} {\bibinfo {author} {\bibfnamefont {J.~M.}\ \bibnamefont
  {Drouin}}, \bibinfo {author} {\bibfnamefont {T.}~\bibnamefont {Chopin}},
  \bibinfo {author} {\bibfnamefont {P.}~\bibnamefont {Nortier}}, \ and\
  \bibinfo {author} {\bibfnamefont {H.}~\bibnamefont {{Van Damme}}},\
  }\href@noop {} {\bibfield  {journal} {\bibinfo  {journal} {J. Colloid
  Interface Sci.}\ }\textbf {\bibinfo {volume} {125}},\ \bibinfo {pages} {314}
  (\bibinfo {year} {1988})}\BibitemShut {NoStop}%
\bibitem [{\citenamefont {Cristiani}\ \emph {et~al.}(2007)\citenamefont
  {Cristiani}, \citenamefont {Grossale},\ and\ \citenamefont
  {Forzatti}}]{Cristiani.2007}%
  \BibitemOpen
  \bibfield  {author} {\bibinfo {author} {\bibfnamefont {C.}~\bibnamefont
  {Cristiani}}, \bibinfo {author} {\bibfnamefont {A.}~\bibnamefont {Grossale}},
  \ and\ \bibinfo {author} {\bibfnamefont {P.}~\bibnamefont {Forzatti}},\
  }\href@noop {} {\bibfield  {journal} {\bibinfo  {journal} {Top. Catal.}\
  }\textbf {\bibinfo {volume} {42-43}},\ \bibinfo {pages} {455} (\bibinfo
  {year} {2007})}\BibitemShut {NoStop}%
\bibitem [{\citenamefont {Sudreau}\ \emph {et~al.}(2022)\citenamefont
  {Sudreau}, \citenamefont {Manneville}, \citenamefont {Servel},\ and\
  \citenamefont {Divoux}}]{Sudreau:2022}%
  \BibitemOpen
  \bibfield  {author} {\bibinfo {author} {\bibfnamefont {I.}~\bibnamefont
  {Sudreau}}, \bibinfo {author} {\bibfnamefont {S.}~\bibnamefont {Manneville}},
  \bibinfo {author} {\bibfnamefont {M.}~\bibnamefont {Servel}}, \ and\ \bibinfo
  {author} {\bibfnamefont {T.}~\bibnamefont {Divoux}},\ }\href@noop {}
  {\bibfield  {journal} {\bibinfo  {journal} {J. Rheol.}\ }\textbf {\bibinfo
  {volume} {66}},\ \bibinfo {pages} {91} (\bibinfo {year} {2022})}\BibitemShut
  {NoStop}%
\bibitem [{\citenamefont {Fauchadour}\ \emph {et~al.}(2002)\citenamefont
  {Fauchadour}, \citenamefont {Kolenda}, \citenamefont {Rouleau}, \citenamefont
  {Barr{\'e}},\ and\ \citenamefont {Normand}}]{Fauchadour.2002}%
  \BibitemOpen
  \bibfield  {author} {\bibinfo {author} {\bibfnamefont {D.}~\bibnamefont
  {Fauchadour}}, \bibinfo {author} {\bibfnamefont {F.}~\bibnamefont {Kolenda}},
  \bibinfo {author} {\bibfnamefont {L.}~\bibnamefont {Rouleau}}, \bibinfo
  {author} {\bibfnamefont {L.}~\bibnamefont {Barr{\'e}}}, \ and\ \bibinfo
  {author} {\bibfnamefont {L.}~\bibnamefont {Normand}},\ }\href@noop {}
  {\bibfield  {journal} {\bibinfo  {journal} {Stud. Surf. Sci. Catal.}\
  }\textbf {\bibinfo {volume} {143}},\ \bibinfo {pages} {453} (\bibinfo {year}
  {2002})}\BibitemShut {NoStop}%
\bibitem [{\citenamefont {de~Rooij}\ \emph {et~al.}(1994)\citenamefont
  {de~Rooij}, \citenamefont {van~den Ende}, \citenamefont {Duits},\ and\
  \citenamefont {Mellema}}]{Rooij:1994}%
  \BibitemOpen
  \bibfield  {author} {\bibinfo {author} {\bibfnamefont {R.}~\bibnamefont
  {de~Rooij}}, \bibinfo {author} {\bibfnamefont {D.}~\bibnamefont {van~den
  Ende}}, \bibinfo {author} {\bibfnamefont {M.~H.~G.}\ \bibnamefont {Duits}}, \
  and\ \bibinfo {author} {\bibfnamefont {J.}~\bibnamefont {Mellema}},\
  }\href@noop {} {\bibfield  {journal} {\bibinfo  {journal} {Phys. Rev. E}\
  }\textbf {\bibinfo {volume} {49}},\ \bibinfo {pages} {3038} (\bibinfo {year}
  {1994})}\BibitemShut {NoStop}%
\bibitem [{\citenamefont {Kao}\ \emph {et~al.}(2021)\citenamefont {Kao},
  \citenamefont {Solomon},\ and\ \citenamefont {Ganesan}}]{Kao.2021}%
  \BibitemOpen
  \bibfield  {author} {\bibinfo {author} {\bibfnamefont {P.-K.}\ \bibnamefont
  {Kao}}, \bibinfo {author} {\bibfnamefont {M.~J.}\ \bibnamefont {Solomon}}, \
  and\ \bibinfo {author} {\bibfnamefont {M.}~\bibnamefont {Ganesan}},\
  }\href@noop {} {\bibfield  {journal} {\bibinfo  {journal} {arXiv:2107.03422}\
  } (\bibinfo {year} {2021})}\BibitemShut {NoStop}%
\bibitem [{\citenamefont {Markutsya}\ \emph {et~al.}(2014)\citenamefont
  {Markutsya}, \citenamefont {Fox},\ and\ \citenamefont
  {Subramaniam}}]{Markutsya:2014}%
  \BibitemOpen
  \bibfield  {author} {\bibinfo {author} {\bibfnamefont {S.}~\bibnamefont
  {Markutsya}}, \bibinfo {author} {\bibfnamefont {R.~O.}\ \bibnamefont {Fox}},
  \ and\ \bibinfo {author} {\bibfnamefont {S.}~\bibnamefont {Subramaniam}},\
  }\href@noop {} {\bibfield  {journal} {\bibinfo  {journal} {Phys. Rev. E}\
  }\textbf {\bibinfo {volume} {89}},\ \bibinfo {pages} {062312} (\bibinfo
  {year} {2014})}\BibitemShut {NoStop}%
\bibitem [{\citenamefont {Boromand}\ \emph {et~al.}(2017)\citenamefont
  {Boromand}, \citenamefont {Jamali},\ and\ \citenamefont
  {Maia}}]{Boromand:2017}%
  \BibitemOpen
  \bibfield  {author} {\bibinfo {author} {\bibfnamefont {A.}~\bibnamefont
  {Boromand}}, \bibinfo {author} {\bibfnamefont {S.}~\bibnamefont {Jamali}}, \
  and\ \bibinfo {author} {\bibfnamefont {J.~M.}\ \bibnamefont {Maia}},\ }\href
  {\doibase 10.1039/C6SM00750C} {\bibfield  {journal} {\bibinfo  {journal}
  {Soft Matter}\ }\textbf {\bibinfo {volume} {13}},\ \bibinfo {pages} {458}
  (\bibinfo {year} {2017})}\BibitemShut {NoStop}%
\bibitem [{\citenamefont {Varga}\ and\ \citenamefont
  {Swan}(2018)}]{Varga:2018}%
  \BibitemOpen
  \bibfield  {author} {\bibinfo {author} {\bibfnamefont {Z.}~\bibnamefont
  {Varga}}\ and\ \bibinfo {author} {\bibfnamefont {J.~W.}\ \bibnamefont
  {Swan}},\ }\href@noop {} {\bibfield  {journal} {\bibinfo  {journal} {J.
  Rheol.}\ }\textbf {\bibinfo {volume} {62}},\ \bibinfo {pages} {405} (\bibinfo
  {year} {2018})}\BibitemShut {NoStop}%
\bibitem [{\citenamefont {Varga}\ \emph {et~al.}(2019)\citenamefont {Varga},
  \citenamefont {Grenard}, \citenamefont {Pecorario}, \citenamefont {Taberlet},
  \citenamefont {Dolique}, \citenamefont {Manneville}, \citenamefont {Divoux},
  \citenamefont {McKinley},\ and\ \citenamefont {Swan}}]{Varga:2019}%
  \BibitemOpen
  \bibfield  {author} {\bibinfo {author} {\bibfnamefont {Z.}~\bibnamefont
  {Varga}}, \bibinfo {author} {\bibfnamefont {V.}~\bibnamefont {Grenard}},
  \bibinfo {author} {\bibfnamefont {S.}~\bibnamefont {Pecorario}}, \bibinfo
  {author} {\bibfnamefont {N.}~\bibnamefont {Taberlet}}, \bibinfo {author}
  {\bibfnamefont {V.}~\bibnamefont {Dolique}}, \bibinfo {author} {\bibfnamefont
  {S.}~\bibnamefont {Manneville}}, \bibinfo {author} {\bibfnamefont
  {T.}~\bibnamefont {Divoux}}, \bibinfo {author} {\bibfnamefont {G.~H.}\
  \bibnamefont {McKinley}}, \ and\ \bibinfo {author} {\bibfnamefont {J.~W.}\
  \bibnamefont {Swan}},\ }\href@noop {} {\bibfield  {journal} {\bibinfo
  {journal} {Proc. Natl. Acad. Sci}\ }\textbf {\bibinfo {volume} {116}},\
  \bibinfo {pages} {12193} (\bibinfo {year} {2019})}\BibitemShut {NoStop}%
\bibitem [{\citenamefont {Derec}\ \emph {et~al.}(2003)\citenamefont {Derec},
  \citenamefont {Ducouret}, \citenamefont {Ajdari},\ and\ \citenamefont
  {Lequeux}}]{Derec:2003}%
  \BibitemOpen
  \bibfield  {author} {\bibinfo {author} {\bibfnamefont {C.}~\bibnamefont
  {Derec}}, \bibinfo {author} {\bibfnamefont {G.}~\bibnamefont {Ducouret}},
  \bibinfo {author} {\bibfnamefont {A.}~\bibnamefont {Ajdari}}, \ and\ \bibinfo
  {author} {\bibfnamefont {F.}~\bibnamefont {Lequeux}},\ }\href@noop {}
  {\bibfield  {journal} {\bibinfo  {journal} {Phys. Rev. E}\ }\textbf {\bibinfo
  {volume} {67}},\ \bibinfo {pages} {061403} (\bibinfo {year}
  {2003})}\BibitemShut {NoStop}%
\bibitem [{\citenamefont {Coussot}\ \emph {et~al.}(2006)\citenamefont
  {Coussot}, \citenamefont {Tabuteau}, \citenamefont {Chateau}, \citenamefont
  {Tocquer},\ and\ \citenamefont {Ovarlez}}]{Coussot:2006}%
  \BibitemOpen
  \bibfield  {author} {\bibinfo {author} {\bibfnamefont {P.}~\bibnamefont
  {Coussot}}, \bibinfo {author} {\bibfnamefont {H.}~\bibnamefont {Tabuteau}},
  \bibinfo {author} {\bibfnamefont {X.}~\bibnamefont {Chateau}}, \bibinfo
  {author} {\bibfnamefont {L.}~\bibnamefont {Tocquer}}, \ and\ \bibinfo
  {author} {\bibfnamefont {G.}~\bibnamefont {Ovarlez}},\ }\href {\doibase
  10.1122/1.2337259} {\bibfield  {journal} {\bibinfo  {journal} {J. Rheol.}\
  }\textbf {\bibinfo {volume} {50}},\ \bibinfo {pages} {975} (\bibinfo {year}
  {2006})}\BibitemShut {NoStop}%
\bibitem [{\citenamefont {Joshi}\ and\ \citenamefont
  {Petekidis}(2018)}]{Joshi.2018}%
  \BibitemOpen
  \bibfield  {author} {\bibinfo {author} {\bibfnamefont {Y.~M.}\ \bibnamefont
  {Joshi}}\ and\ \bibinfo {author} {\bibfnamefont {G.}~\bibnamefont
  {Petekidis}},\ }\href@noop {} {\bibfield  {journal} {\bibinfo  {journal}
  {Rheol. Acta}\ }\textbf {\bibinfo {volume} {57}},\ \bibinfo {pages} {521}
  (\bibinfo {year} {2018})}\BibitemShut {NoStop}%
\bibitem [{\citenamefont {Bouchaud}(1992)}]{Bouchaud:1992}%
  \BibitemOpen
  \bibfield  {author} {\bibinfo {author} {\bibfnamefont {J.-P.}\ \bibnamefont
  {Bouchaud}},\ }\href@noop {} {\bibfield  {journal} {\bibinfo  {journal} {J.
  Phys. I. France}\ }\textbf {\bibinfo {volume} {2}},\ \bibinfo {pages} {1705}
  (\bibinfo {year} {1992})}\BibitemShut {NoStop}%
\bibitem [{\citenamefont {Berthier}(2003)}]{Berthier:2003}%
  \BibitemOpen
  \bibfield  {author} {\bibinfo {author} {\bibfnamefont {L.}~\bibnamefont
  {Berthier}},\ }\href {\doibase 10.1088/0953-8984/15/11/317} {\bibfield
  {journal} {\bibinfo  {journal} {J. Phys.: Condens. Matter}\ }\textbf
  {\bibinfo {volume} {15}},\ \bibinfo {pages} {S933} (\bibinfo {year}
  {2003})}\BibitemShut {NoStop}%
\bibitem [{\citenamefont {Viasnoff}\ and\ \citenamefont
  {Lequeux}(2002)}]{Viasnoff:2002}%
  \BibitemOpen
  \bibfield  {author} {\bibinfo {author} {\bibfnamefont {V.}~\bibnamefont
  {Viasnoff}}\ and\ \bibinfo {author} {\bibfnamefont {F.}~\bibnamefont
  {Lequeux}},\ }\href@noop {} {\bibfield  {journal} {\bibinfo  {journal} {Phys.
  Rev. Lett.}\ }\textbf {\bibinfo {volume} {89}},\ \bibinfo {pages} {065701}
  (\bibinfo {year} {2002})}\BibitemShut {NoStop}%
\bibitem [{\citenamefont {Lacks}\ and\ \citenamefont
  {Osborne}(2004)}]{Lacks:2004}%
  \BibitemOpen
  \bibfield  {author} {\bibinfo {author} {\bibfnamefont {D.~J.}\ \bibnamefont
  {Lacks}}\ and\ \bibinfo {author} {\bibfnamefont {M.~J.}\ \bibnamefont
  {Osborne}},\ }\href@noop {} {\bibfield  {journal} {\bibinfo  {journal} {Phys.
  Rev. Lett.}\ }\textbf {\bibinfo {volume} {93}},\ \bibinfo {pages} {255501}
  (\bibinfo {year} {2004})}\BibitemShut {NoStop}%
\bibitem [{\citenamefont {Joshi}(2014)}]{Joshi:2014}%
  \BibitemOpen
  \bibfield  {author} {\bibinfo {author} {\bibfnamefont {Y.~M.}\ \bibnamefont
  {Joshi}},\ }\href@noop {} {\bibfield  {journal} {\bibinfo  {journal} {Annu.
  Rev. Chem. Biomol. Eng.}\ }\textbf {\bibinfo {volume} {5}},\ \bibinfo {pages}
  {181} (\bibinfo {year} {2014})}\BibitemShut {NoStop}%
\bibitem [{\citenamefont {Pashine}\ \emph {et~al.}(2019)\citenamefont
  {Pashine}, \citenamefont {Hexner}, \citenamefont {Liu},\ and\ \citenamefont
  {Nagel}}]{Pashine:2019}%
  \BibitemOpen
  \bibfield  {author} {\bibinfo {author} {\bibfnamefont {N.}~\bibnamefont
  {Pashine}}, \bibinfo {author} {\bibfnamefont {D.}~\bibnamefont {Hexner}},
  \bibinfo {author} {\bibfnamefont {A.~J.}\ \bibnamefont {Liu}}, \ and\
  \bibinfo {author} {\bibfnamefont {S.~R.}\ \bibnamefont {Nagel}},\ }\href@noop
  {} {\bibfield  {journal} {\bibinfo  {journal} {Sci. Adv.}\ }\textbf {\bibinfo
  {volume} {5}} (\bibinfo {year} {2019})}\BibitemShut {NoStop}%
\bibitem [{\citenamefont {Hexner}\ \emph {et~al.}(2020)\citenamefont {Hexner},
  \citenamefont {Pashine}, \citenamefont {Liu},\ and\ \citenamefont
  {Nagel}}]{Hexner:2020}%
  \BibitemOpen
  \bibfield  {author} {\bibinfo {author} {\bibfnamefont {D.}~\bibnamefont
  {Hexner}}, \bibinfo {author} {\bibfnamefont {N.}~\bibnamefont {Pashine}},
  \bibinfo {author} {\bibfnamefont {A.~J.}\ \bibnamefont {Liu}}, \ and\
  \bibinfo {author} {\bibfnamefont {S.~R.}\ \bibnamefont {Nagel}},\ }\href@noop
  {} {\bibfield  {journal} {\bibinfo  {journal} {Phys. Rev. Research}\ }\textbf
  {\bibinfo {volume} {2}},\ \bibinfo {pages} {043231} (\bibinfo {year}
  {2020})}\BibitemShut {NoStop}%
\bibitem [{\citenamefont {Johnson}\ \emph {et~al.}(2019)\citenamefont
  {Johnson}, \citenamefont {Zia}, \citenamefont {Moghimi},\ and\ \citenamefont
  {Petekidis}}]{Johnson:2019}%
  \BibitemOpen
  \bibfield  {author} {\bibinfo {author} {\bibfnamefont {L.~C.}\ \bibnamefont
  {Johnson}}, \bibinfo {author} {\bibfnamefont {R.~N.}\ \bibnamefont {Zia}},
  \bibinfo {author} {\bibfnamefont {E.}~\bibnamefont {Moghimi}}, \ and\
  \bibinfo {author} {\bibfnamefont {G.}~\bibnamefont {Petekidis}},\ }\href@noop
  {} {\bibfield  {journal} {\bibinfo  {journal} {J. Rheol.}\ }\textbf {\bibinfo
  {volume} {63}},\ \bibinfo {pages} {583} (\bibinfo {year} {2019})}\BibitemShut
  {NoStop}%
\bibitem [{\citenamefont {Keshavarz}\ \emph {et~al.}(2021)\citenamefont
  {Keshavarz}, \citenamefont {Gomes~Rodrigues}, \citenamefont {Champenois},
  \citenamefont {Frith}, \citenamefont {Ilavsky}, \citenamefont {Geri},
  \citenamefont {Divoux}, \citenamefont {McKinley},\ and\ \citenamefont
  {Poulesquen}}]{Keshavarz:2021}%
  \BibitemOpen
  \bibfield  {author} {\bibinfo {author} {\bibfnamefont {B.}~\bibnamefont
  {Keshavarz}}, \bibinfo {author} {\bibfnamefont {D.}~\bibnamefont
  {Gomes~Rodrigues}}, \bibinfo {author} {\bibfnamefont {J.-B.}\ \bibnamefont
  {Champenois}}, \bibinfo {author} {\bibfnamefont {M.~G.}\ \bibnamefont
  {Frith}}, \bibinfo {author} {\bibfnamefont {J.}~\bibnamefont {Ilavsky}},
  \bibinfo {author} {\bibfnamefont {M.}~\bibnamefont {Geri}}, \bibinfo {author}
  {\bibfnamefont {T.}~\bibnamefont {Divoux}}, \bibinfo {author} {\bibfnamefont
  {G.~H.}\ \bibnamefont {McKinley}}, \ and\ \bibinfo {author} {\bibfnamefont
  {A.}~\bibnamefont {Poulesquen}},\ }\href@noop {} {\bibfield  {journal}
  {\bibinfo  {journal} {Proc. Natl. Acad. Sci. U.S.A.}\ }\textbf {\bibinfo
  {volume} {118}} (\bibinfo {year} {2021})}\BibitemShut {NoStop}%
\bibitem [{\citenamefont {Mohraz}\ and\ \citenamefont
  {Solomon}(2005)}]{Mohraz:2005}%
  \BibitemOpen
  \bibfield  {author} {\bibinfo {author} {\bibfnamefont {A.}~\bibnamefont
  {Mohraz}}\ and\ \bibinfo {author} {\bibfnamefont {M.~J.}\ \bibnamefont
  {Solomon}},\ }\href@noop {} {\bibfield  {journal} {\bibinfo  {journal} {J.
  Rheol.}\ }\textbf {\bibinfo {volume} {49}},\ \bibinfo {pages} {657} (\bibinfo
  {year} {2005})}\BibitemShut {NoStop}%
\bibitem [{\citenamefont {Rajaram}\ and\ \citenamefont
  {Mohraz}(2010)}]{Rajaram:2010}%
  \BibitemOpen
  \bibfield  {author} {\bibinfo {author} {\bibfnamefont {B.}~\bibnamefont
  {Rajaram}}\ and\ \bibinfo {author} {\bibfnamefont {A.}~\bibnamefont
  {Mohraz}},\ }\href@noop {} {\bibfield  {journal} {\bibinfo  {journal} {Soft
  Matter}\ }\textbf {\bibinfo {volume} {6}},\ \bibinfo {pages} {2246} (\bibinfo
  {year} {2010})}\BibitemShut {NoStop}%
\bibitem [{\citenamefont {Santos}\ \emph {et~al.}(2013)\citenamefont {Santos},
  \citenamefont {Campanella},\ and\ \citenamefont {Carignano}}]{Santos:2013}%
  \BibitemOpen
  \bibfield  {author} {\bibinfo {author} {\bibfnamefont {P.~H.~S.}\
  \bibnamefont {Santos}}, \bibinfo {author} {\bibfnamefont {O.~H.}\
  \bibnamefont {Campanella}}, \ and\ \bibinfo {author} {\bibfnamefont {M.~A.}\
  \bibnamefont {Carignano}},\ }\href@noop {} {\bibfield  {journal} {\bibinfo
  {journal} {Soft Matter}\ }\textbf {\bibinfo {volume} {9}},\ \bibinfo {pages}
  {709} (\bibinfo {year} {2013})}\BibitemShut {NoStop}%
\bibitem [{\citenamefont {van Doorn}\ \emph {et~al.}(2018)\citenamefont {van
  Doorn}, \citenamefont {Verweij}, \citenamefont {Sprakel},\ and\ \citenamefont
  {van~der Gucht}}]{Doorn:2018}%
  \BibitemOpen
  \bibfield  {author} {\bibinfo {author} {\bibfnamefont {J.~M.}\ \bibnamefont
  {van Doorn}}, \bibinfo {author} {\bibfnamefont {J.~E.}\ \bibnamefont
  {Verweij}}, \bibinfo {author} {\bibfnamefont {J.}~\bibnamefont {Sprakel}}, \
  and\ \bibinfo {author} {\bibfnamefont {J.}~\bibnamefont {van~der Gucht}},\
  }\href@noop {} {\bibfield  {journal} {\bibinfo  {journal} {Phys. Rev. Lett.}\
  }\textbf {\bibinfo {volume} {120}},\ \bibinfo {pages} {208005} (\bibinfo
  {year} {2018})}\BibitemShut {NoStop}%
\bibitem [{\citenamefont {Johnson}\ \emph {et~al.}(2018)\citenamefont
  {Johnson}, \citenamefont {Landrum},\ and\ \citenamefont
  {Zia}}]{Johnson:2018}%
  \BibitemOpen
  \bibfield  {author} {\bibinfo {author} {\bibfnamefont {L.~C.}\ \bibnamefont
  {Johnson}}, \bibinfo {author} {\bibfnamefont {B.~J.}\ \bibnamefont
  {Landrum}}, \ and\ \bibinfo {author} {\bibfnamefont {R.~N.}\ \bibnamefont
  {Zia}},\ }\href@noop {} {\bibfield  {journal} {\bibinfo  {journal} {Soft
  Matter}\ }\textbf {\bibinfo {volume} {14}},\ \bibinfo {pages} {5048}
  (\bibinfo {year} {2018})}\BibitemShut {NoStop}%
\bibitem [{\citenamefont {Murphy}\ \emph {et~al.}(2020)\citenamefont {Murphy},
  \citenamefont {Kruppe},\ and\ \citenamefont {Jaeger}}]{Murphy:2020}%
  \BibitemOpen
  \bibfield  {author} {\bibinfo {author} {\bibfnamefont {K.~A.}\ \bibnamefont
  {Murphy}}, \bibinfo {author} {\bibfnamefont {J.~W.}\ \bibnamefont {Kruppe}},
  \ and\ \bibinfo {author} {\bibfnamefont {H.~M.}\ \bibnamefont {Jaeger}},\
  }\href@noop {} {\bibfield  {journal} {\bibinfo  {journal} {Phys. Rev. Lett.}\
  }\textbf {\bibinfo {volume} {124}},\ \bibinfo {pages} {168002} (\bibinfo
  {year} {2020})}\BibitemShut {NoStop}%
\bibitem [{\citenamefont {Mandal}\ \emph {et~al.}(2021)\citenamefont {Mandal},
  \citenamefont {Tapias},\ and\ \citenamefont {Sollich}}]{Mandal:2021}%
  \BibitemOpen
  \bibfield  {author} {\bibinfo {author} {\bibfnamefont {R.}~\bibnamefont
  {Mandal}}, \bibinfo {author} {\bibfnamefont {D.}~\bibnamefont {Tapias}}, \
  and\ \bibinfo {author} {\bibfnamefont {P.}~\bibnamefont {Sollich}},\
  }\href@noop {} {\bibfield  {journal} {\bibinfo  {journal} {Phys. Res.
  Research}\ }\textbf {\bibinfo {volume} {3}},\ \bibinfo {pages} {043153}
  (\bibinfo {year} {2021})}\BibitemShut {NoStop}%
\bibitem [{\citenamefont {Hendricks}\ \emph {et~al.}(2019)\citenamefont
  {Hendricks}, \citenamefont {Louhichi}, \citenamefont {Metri}, \citenamefont
  {Fournier}, \citenamefont {Reddy}, \citenamefont {Bouteiller}, \citenamefont
  {Cloitre}, \citenamefont {Clasen}, \citenamefont {Vlassopoulos},\ and\
  \citenamefont {Briels}}]{Hendricks:2019}%
  \BibitemOpen
  \bibfield  {author} {\bibinfo {author} {\bibfnamefont {J.}~\bibnamefont
  {Hendricks}}, \bibinfo {author} {\bibfnamefont {A.}~\bibnamefont {Louhichi}},
  \bibinfo {author} {\bibfnamefont {V.}~\bibnamefont {Metri}}, \bibinfo
  {author} {\bibfnamefont {R.}~\bibnamefont {Fournier}}, \bibinfo {author}
  {\bibfnamefont {N.}~\bibnamefont {Reddy}}, \bibinfo {author} {\bibfnamefont
  {L.}~\bibnamefont {Bouteiller}}, \bibinfo {author} {\bibfnamefont
  {M.}~\bibnamefont {Cloitre}}, \bibinfo {author} {\bibfnamefont
  {C.}~\bibnamefont {Clasen}}, \bibinfo {author} {\bibfnamefont
  {D.}~\bibnamefont {Vlassopoulos}}, \ and\ \bibinfo {author} {\bibfnamefont
  {W.~J.}\ \bibnamefont {Briels}},\ }\href@noop {} {\bibfield  {journal}
  {\bibinfo  {journal} {Phys. Rev. Lett.}\ }\textbf {\bibinfo {volume} {123}},\
  \bibinfo {pages} {218003} (\bibinfo {year} {2019})}\BibitemShut {NoStop}%
\bibitem [{\citenamefont {Zia}\ and\ \citenamefont {Brady}(2013)}]{Zia:2013}%
  \BibitemOpen
  \bibfield  {author} {\bibinfo {author} {\bibfnamefont {R.~N.}\ \bibnamefont
  {Zia}}\ and\ \bibinfo {author} {\bibfnamefont {J.~F.}\ \bibnamefont
  {Brady}},\ }\href@noop {} {\bibfield  {journal} {\bibinfo  {journal} {J.
  Rheol.}\ }\textbf {\bibinfo {volume} {57}},\ \bibinfo {pages} {457} (\bibinfo
  {year} {2013})}\BibitemShut {NoStop}%
\bibitem [{\citenamefont {Bouzid}\ and\ \citenamefont
  {Gado}(2020)}]{Bouzid:2020}%
  \BibitemOpen
  \bibfield  {author} {\bibinfo {author} {\bibfnamefont {M.}~\bibnamefont
  {Bouzid}}\ and\ \bibinfo {author} {\bibfnamefont {E.~D.}\ \bibnamefont
  {Gado}},\ }\enquote {\bibinfo {title} {Mechanics of soft gels: Linear and
  nonlinear response},}\ in\ \href@noop {} {\emph {\bibinfo {booktitle}
  {Handbook of Materials Modeling: Applications: Current and Emerging
  Materials}}}\ (\bibinfo  {publisher} {Springer International Publishing},\
  \bibinfo {year} {2020})\ pp.\ \bibinfo {pages} {1719--1746}\BibitemShut
  {NoStop}%
\end{thebibliography}

%


\clearpage
\newpage
\onecolumngrid
\setcounter{page}{1}
\setcounter{figure}{0}
\global\def\thefigure{S\arabic{figure}}
\setcounter{table}{0}
\global\def\thetable{S\arabic{table}}
\setcounter{equation}{0}
\global\def\theequation{S\arabic{equation}}

\begin{center}
    {\large\bf Residual stresses and shear-induced overaging in boehmite gels}
\end{center}

\begin{center}
    {\large\bf{\sc Supplemental Material}}
\end{center}


\section{Full temporal evolution of rheological observables}

Figure~S1(a) displays the temporal evolution of the shear rate $\gp$ effectively applied (as measured by the rheometer) during the flow cessation step and normalized by $\gpp$. In all cases, it takes about 0.5~s for the shear rate to reach $\gp/\gpp=10^{-3}$. Therefore, we may assume the shear rate to be effectively zero for $t>1$~s, which is the reason why we report the stress response starting at $t=1$~s in Fig.~1 in the main text. For the sake of completeness, we also report in Fig.~S1(b) the stress responses for $t>0.5$~ms.

\begin{figure}[h!]
    \centering
    \includegraphics[width=0.91\linewidth]{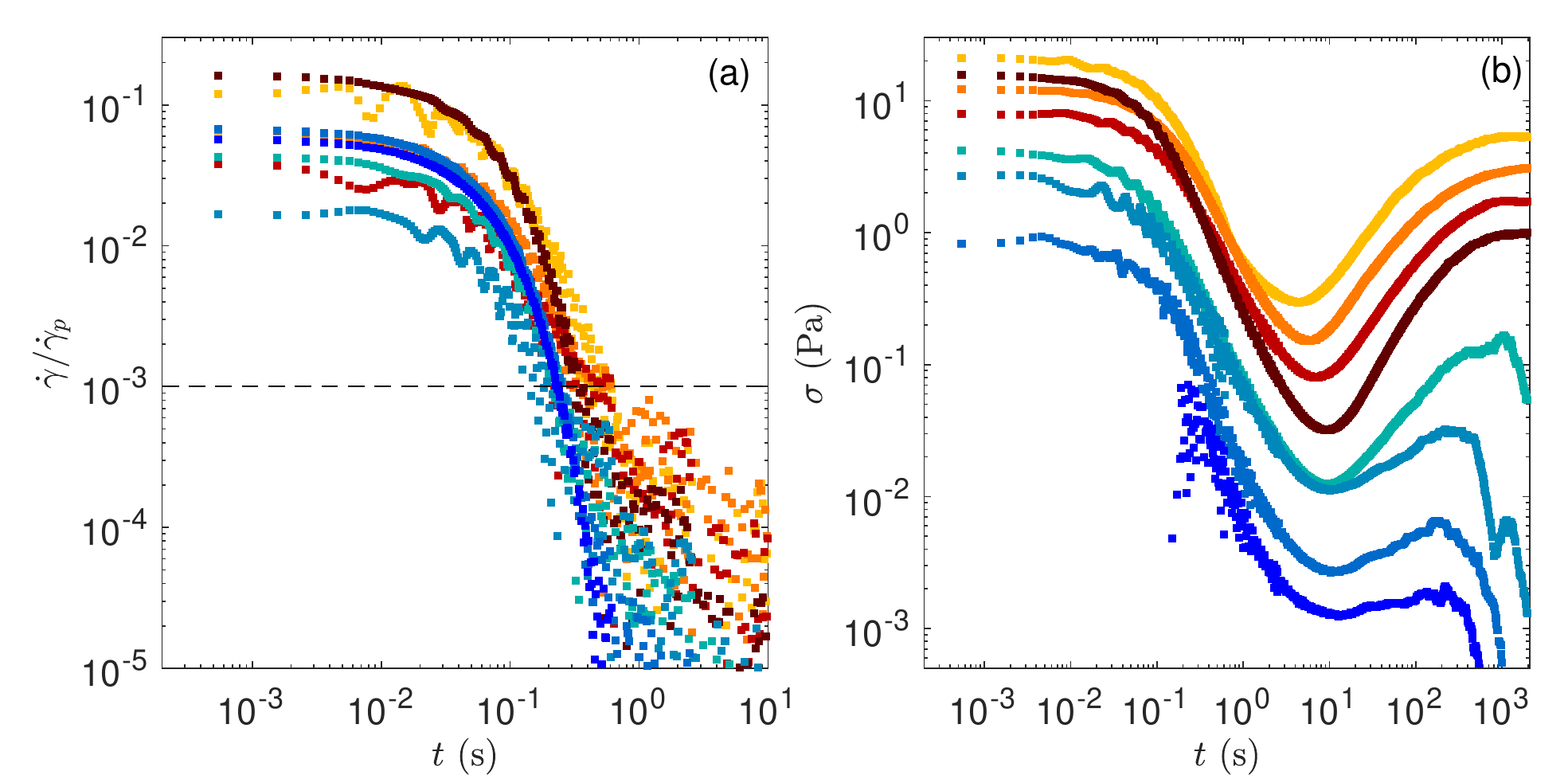}
    \caption{\label{Figure9} (a) Temporal evolution of the shear rate $\gp$ effectively applied and normalized by the shear rate $\gpp$ imposed prior to the flow cessation step. The black horizontal dashed line highlights $\gp/\gpp=10^{-3}$. (b)~Stress response $\sigma(t)$ of a boehmite gel following flow cessation induced by imposing a decreasing step of shear rate from $\gp=\gpp$ to $\gp=0$ at $t=0$~s. The colors code for $\gpp=3$~s$^{-1}$(\textcolor{color3}{$\blacksquare$}), 5~s$^{-1}$(\textcolor{color5}{$\blacksquare$}), 10~s$^{-1}$(\textcolor{color10}{$\blacksquare$}), 15~s$^{-1}$(\textcolor{color15}{$\blacksquare$}), 40~s$^{-1}$(\textcolor{color40}{$\blacksquare$}), 50~s$^{-1}$(\textcolor{color50}{$\blacksquare$}), 200~s$^{-1}$(\textcolor{color200}{$\blacksquare$}), and 500~s$^{-1}$(\textcolor{color500}{$\blacksquare$}).
    } 
\end{figure}

\newpage
\section{Robustness of non-monotonic stress relaxation}

Figure~S2 reports additional stress responses following flow cessation as described in the main text. Using the same Couette geometry and the same rheometer as in the main text, we checked that the stress has reached a stationary value by monitoring $\sigma(t)$ over $10^{4}$~s instead of 2000~s (see square symbols). Moreover, measurements performed over 2000~s using a sandblasted cone-and-plate geometry connected to a different stress-controlled rheometer (MCR 302, Anton Paar) display a very similar non-monotonic evolution and quantitatively point to the same residual stresses (see dotted lines for $\gp=2$ and 5~s$^{-1}$). These measurements emphasize the robustness of the anomalous stress relaxation towards significant residual stresses for $\gpp<\gpc$. {Note that the stress relaxations do not strictly coincide in the two different geometries, which is not surprising as the details of the global stress dynamics following flow cessation must depend on the local deformation of the microstructure, which is itself likely to depend on the (geometry-dependent) local shear rate and shear stress fields.}

\begin{figure}[h!]
    \centering
    \includegraphics[width=0.5\linewidth]{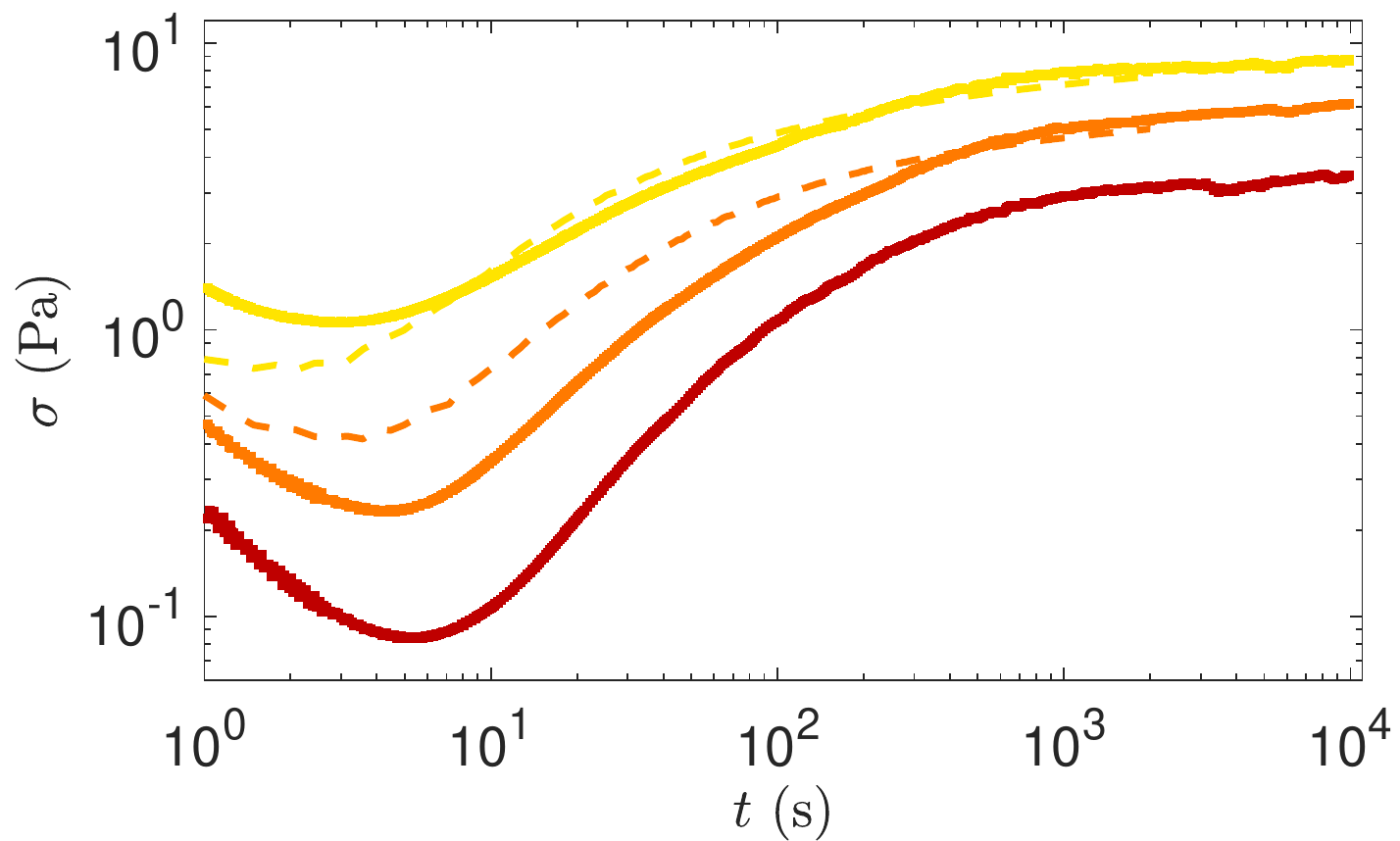}
    \caption{\label{Figure5} Stress response $\sigma(t)$ of a boehmite gel following flow cessation induced by imposing a decreasing step of shear rate from $\gp=\gpp$ to $\gp=0$ at $t=0$~s. Square symbols correspond to experiments performed in the same Couette geometry as in the main text but over a longer duration of $10^4$~s, while dotted lines refer to experiments performed over 2000~s in a sandblasted cone-and-plate geometry (angle of $2^\circ$ and diameter of 40~mm) connected to an MCR 302 (Anton Paar) stress-controlled rheometer. The colors code for $\gpp=2$~s$^{-1}$ (\textcolor{color2}{$\blacksquare$}), 5~s$^{-1}$(\textcolor{color5}{$\blacksquare$}), and 10~s$^{-1}$(\textcolor{color10}{$\blacksquare$}). The duration of the shear step at $\gpp$ is $t_p=600$~s in all cases.} 
\end{figure}

\newpage
\section{Complementary analysis of the stress responses}

In Fig.~S3, we show complementary observables extracted from Fig.~1 in the main text, as a function of the shear rate $\gpp$, namely the time $t_{\min}$ at which the stress reaches a minimum [Fig.~S3(a)], and the loss factor $\tan \delta$ [Fig.~S3(b)]. Both observables display a dual behavior depending on the position of $\gpp$ relative to $\gpc$. For $\gpp<\gpc$, the data are well fitted by a logarithmic dependence in $\gpp$, while they become independent of $\gpp$ above $\gpc$. Such a behavior is fully consistent with the trends observed in Fig.~2 in the main text for the residual stress and the viscoelastic moduli. Note that the stress minimum becomes hardly measurable after strong shearing, as the stress response flattens with $\gpp$, hence the large error bars on $t_{\min}$ for the highest values of $\gpp$. Moreover, the fact that $\tan\delta$ increases with decreasing $\gpp$ indicates that the dissipative component increases more steeply than the elastic component as $\gpp$ decreases. 

\begin{figure*}[h!]
    \centering
    \includegraphics[width=0.8\linewidth]{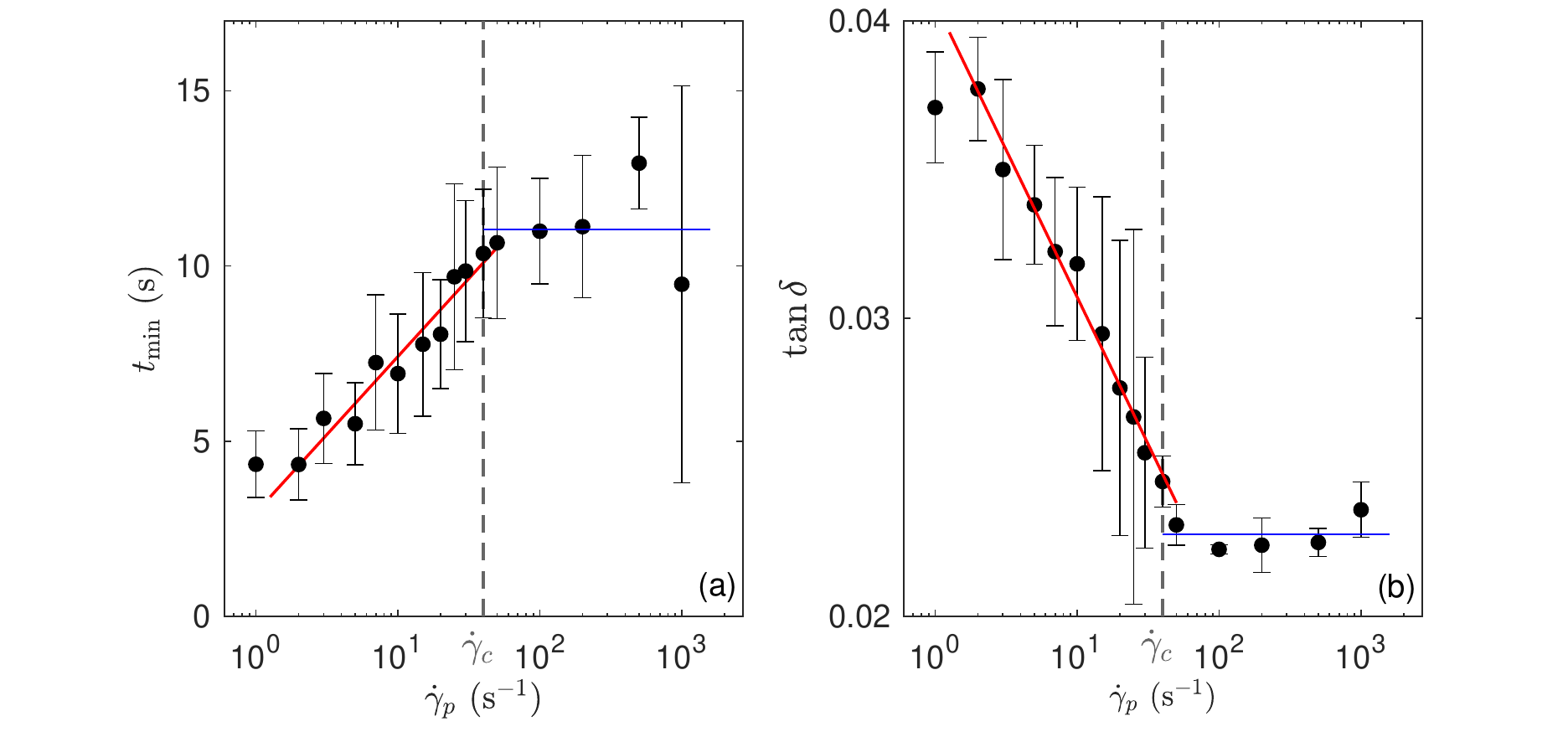} 
    \caption{\label{Figure7} (a) Time $t_{\min}$ at which the stress reaches a minimum and (b)~loss factor $\tan \delta$ vs.~ the shear rate $\dot \gamma_p$ applied prior to flow cessation. The vertical dashed line highlights the critical shear rate $\gpc= 40$ s$^{-1}$ below which the sample grows a residual stress. Blue lines highlight $t_{\min}=11 \pm 1$~s and $\tan\delta = 0.0228 \pm 0.0005$, and red lines are the best logarithmic fits of the data for $\gpp<\gpc$: $X=a_X\log\gpp+b_X$ with $a_t=4.45$~s, $a_\delta=-0.01$, $b_t=2.96$~s, and $b_\delta=0.04$.} 
\end{figure*}

\newpage
\section{Influence of the shearing duration on the stress relaxation}

Figure~S4 presents the stress responses measured after flow cessation following shearing at $\gpp$ over different durations $t_{p}$. It reveals that the accumulated strain $\gamma_p=\gpp t_p$ has a crucial influence on the stress relaxation, and thus on the subsequent residual stress, as further analyzed in Fig.~3 in the main text. 

\begin{figure}[h!]
    \centering
    \includegraphics[width=0.6\linewidth]{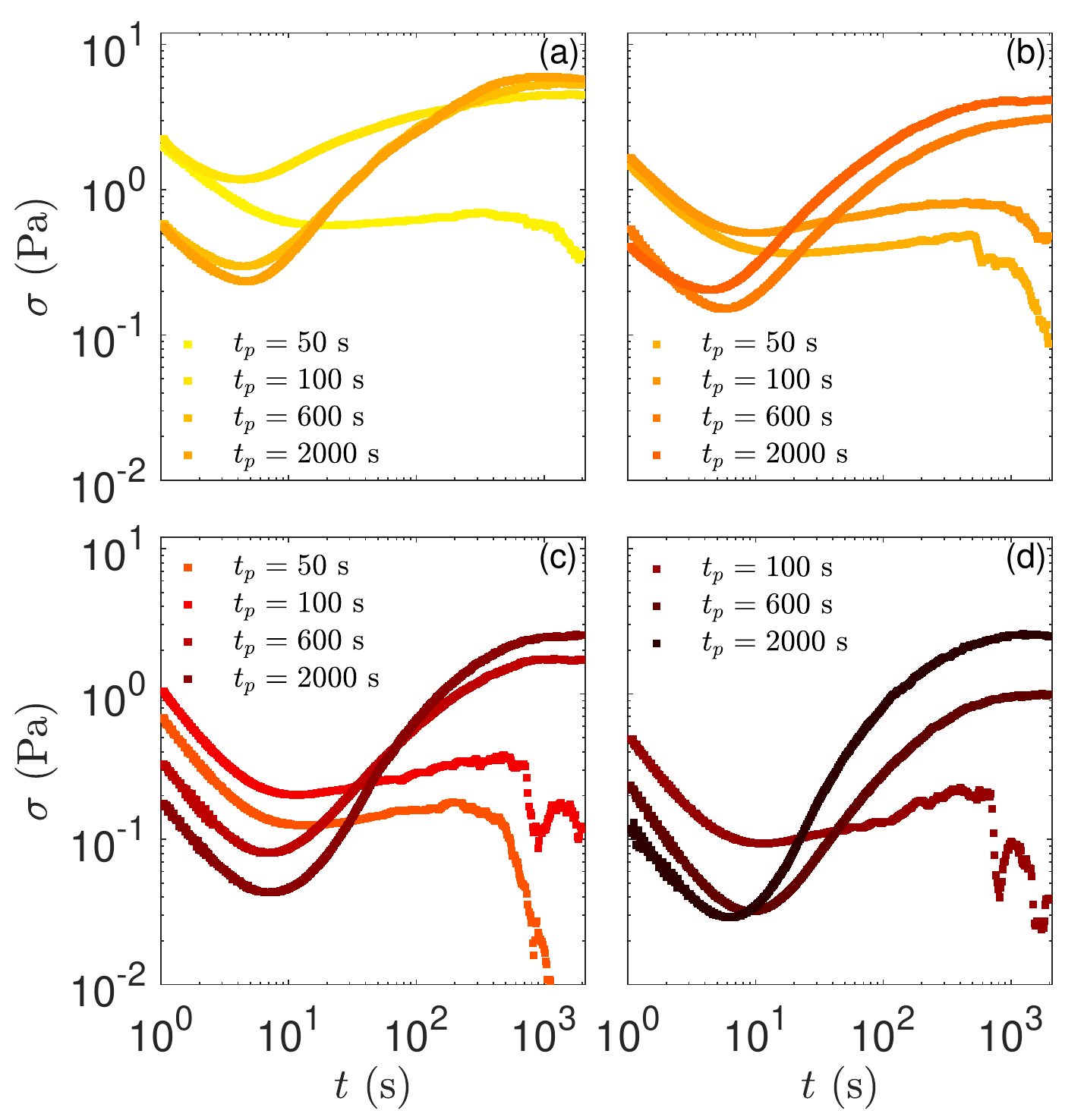}
    \caption{\label{Figure6} Stress response $\sigma(t)$ of a boehmite gel following a flow cessation induced by imposing a shear step from $\gp=\gpp$ to $\gp=0$ at $t=0$~s.  The colors code for (a)~$\gpp=3$~s$^{-1}$, (b)~ $\gpp=5$~s$^{-1}$, (c)~$\gpp=10$~s$^{-1}$, and (d)~$\gpp=15$~s$^{-1}$, and for different durations $t_{p}$ as indicated in the graphs.}
\end{figure}

\newpage
\section{Master curve for the loss factor vs. the residual stress}

From the master curves for $G'$ and $G''$ vs. $\sigma_{\rm res}$ and the linear fits shown in Fig.~4 in the main text, it is straightforward that the loss tangent $\tan\delta=G''/G'$ should also follow a master curve as a function of the residual stress, independent of the shear rate $\gpp$ and duration $t_p$. This is checked in Fig.~S5, where the ratio of the fits for $G''$ and $G'$ (see black line) is shown to account well for the evolution of the loss tangent with $\sigma_{\rm res}$.

\begin{figure}[h!]
    \centering
    \includegraphics[width=0.5\linewidth]{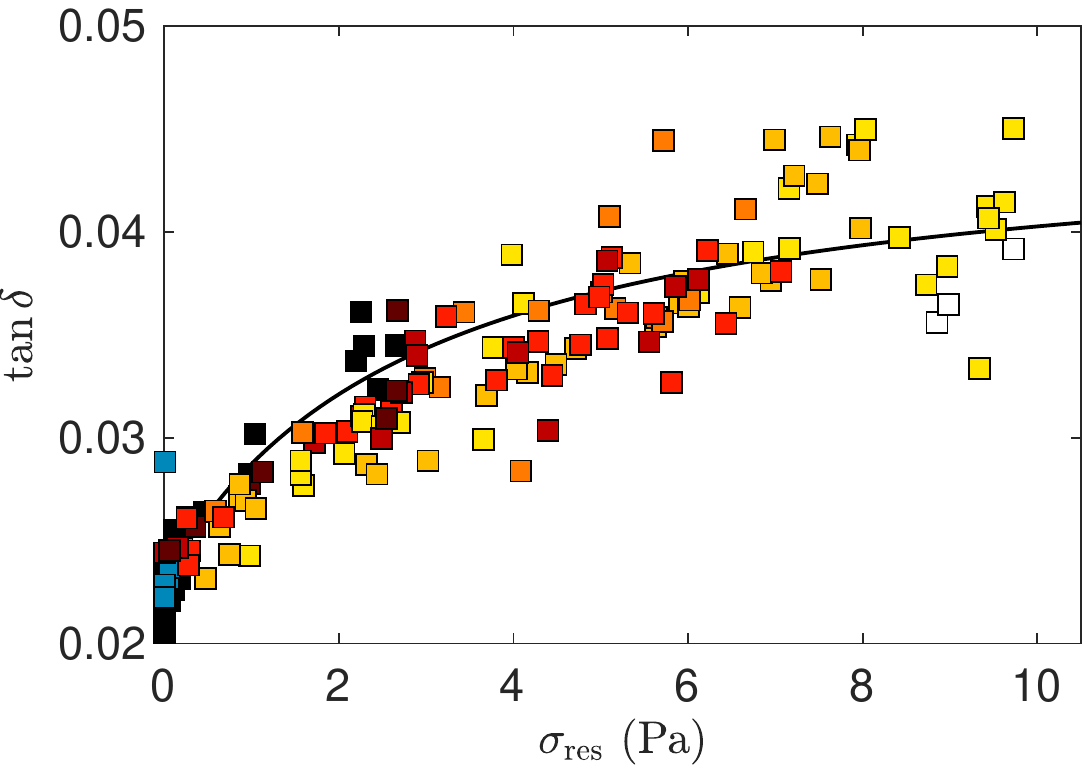}
    \caption{\label{Figure8} Loss factor $\tan \delta$ vs.~residual stress $\sigma_{\rm res}$ for a boehmite gel submitted to various shear histories. Colors code for $\gpp=2$~s$^{-1}$ (\textcolor{color2}{$\blacksquare$}), 3~s$^{-1}$(\textcolor{color3}{$\blacksquare$}), 5~s$^{-1}$(\textcolor{color5}{$\blacksquare$}), 7~s$^{-1}$(\textcolor{color7}{$\blacksquare$}), 10~s$^{-1}$(\textcolor{color10}{$\blacksquare$}), 15~s$^{-1}$(\textcolor{color15}{$\blacksquare$}), and 50~s$^{-1}$(\textcolor{color50}{$\blacksquare$}) imposed over various durations $t_p$ in independent experiments. The black line corresponds to the ratio of the linear fits of $G''$ and $G'$ vs. $\sigma_{\rm res}$ shown in Fig.~4 in the main text, namely $G''_0(1+\lambda'' \sigma_{\rm res})/[G'_0 (1+\lambda' \sigma_{\rm res})]$, with the fit parameters given in the main text.} 
\end{figure}

\end{document}